\documentclass[preprint2]{aastex}
\usepackage{graphicx,natbib,slashbox,multirow}
\usepackage{url}
\usepackage{color}

\begin{document}   

\title{The SDSS Coadd: A Galaxy Photometric Redshift Catalog}      

\author{
Ribamar R. R. Reis$^{1,2}$,
Marcelle Soares-Santos$^{1,3}$,
James Annis$^{1}$,
Scott Dodelson$^{1,4,5}$,
Jiangang Hao$^{1}$,
David Johnston$^{1}$,
Jeffrey Kubo$^{1}$,
Huan Lin$^{1}$,
Hee-Jong Seo$^{6}$,
Melanie Simet$^{4}$
}

\affil{
${}^{1}$Center for Particle Astrophysics, Fermi National Accelerator Laboratory, Batavia, IL 60510 \\
${}^{2}$Instituto de F\'\i sica, Universidade Federal do Rio de Janeiro, CEP 21941-972, Rio de Janeiro, RJ, Brazil \\
${}^{3}$Instituto de Astronomia, Geof\'\i sica e Ciencias Atmosf\'ericas, Universidade de S\~ao Paulo, CEP 05508-090, S\~ao Paulo, SP, Brazil \\
${}^{4}$Department of Astronomy \& Astrophysics,
The University of Chicago, Chicago, IL 60637 \\
${}^{5}$Kavli Institute for Cosmological Physics, Chicago, IL 60637 \\
${}^{6}$Berkeley Center for Cosmological Physics, LBL and Department of
Physics, University of California, Berkeley, CA, USA 94720 \\
}



\begin{abstract}

We present and describe a catalog of galaxy photometric redshifts (photo-z's) 
for the  Sloan Digital Sky Survey (SDSS) Coadd Data. 
We use the Artificial Neural Network (ANN) technique to calculate photo-z's 
and the Nearest Neighbor Error (NNE) method to estimate photo-z errors for 
$\sim$ 13 million objects classified as galaxies in the coadd with $r < 24.5$.  
The photo-z and photo-z error estimators are trained and validated on a 
sample of $\sim 83,000$ galaxies that have SDSS photometry and 
spectroscopic redshifts measured by the SDSS Data Release 7 (DR7), 
the Canadian Network for Observational Cosmology Field Galaxy Survey (CNOC2), 
the Deep Extragalactic Evolutionary Probe Data Release 3(DEEP2 DR3), 
the VIsible imaging Multi-Object Spectrograph - Very Large Telescope Deep Survey 
(VVDS) and the WiggleZ Dark Energy Survey. 
For the best ANN methods we have tried,
we find that 68\% of the galaxies in the validation set have a photo-z
error smaller than
$\sigma_{68} =0.031$.
After presenting our results and quality tests, we provide a short guide
for users accessing the public data.

\end{abstract}

\keywords{photometric redshifts sdss -- Sloan Digital Sky Survey} 


\section{Introduction}\label{int}

In recent years, digital sky surveys obtained multi-band imaging
for of order a hundred million galaxies, however we have spectroscopic
redshifts available for only over one million galaxies. Deep, wide-area
surveys planned for the next decades will increase the number of 
galaxies with multi-band photometry to a few billion and we will
only be able to obtain spectroscopic redshifts for a small fraction
of these objects, due to technological and financial limitations.
As a result, substantial effort has been going into 
developing photometric redshift (photo-z) techniques, which use 
multi-band photometry to estimate approximate galaxy redshifts. For many 
applications in extragalactic astronomy and cosmology, the resulting 
photometric redshift precision is sufficient for the science goals at 
hand, provided one can accurately characterize the uncertainties in the 
photo-z estimates. 

Two broad categories of photo-z estimators are in wide use: 
template-fitting and training set methods. In template-fitting, one 
assigns a redshift to a galaxy by finding 
the redshifted spectral energy distribution (SED), selected 
from a library of templates, 
that best reproduces the observed fluxes in the broadband filters.
By contrast, in the training set approach, one 
uses a training set of galaxies with 
spectroscopic redshifts and photometry to derive an empirical relation
between photometric observables (e.g., magnitudes, colors, and morphological 
indicators) and redshift. 
Examples of empirical methods include Polynomial Fitting \citep{con95b}, 
the Nearest Neighbor method \citep{csa03}, 
the Nearest Neighbor Polynomial (NNP) technique \citep{oya08}, 
Artificial Neural Networks (ANN) \citep{col04,van04,dab07}, and 
Support Vector Machines \citep{wad04}. When a large spectroscopic 
training set that is representative of the photometric data set to be 
analyzed is 
available, training set techniques typically outperform template-fitting 
methods, in the sense that the photo-z estimates have smaller scatter 
and bias with respect to the true redshifts \citep{oya08}. On the 
other hand, template-fitting can be applied to a photometric sample 
for which relatively few spectroscopic analogs exist.
For a comprehensive review and comparison of photo-z methods,  
see \cite{oya08}.

In this paper, we present a publicly available galaxy photometric redshift 
catalog for the coadd data which is part of the Seventh Data Release (DR7) 
of the Sloan Digital Sky 
Survey (SDSS) imaging catalog \citep{bla03b,eis01,gun98,ive04,str02,yor00,aba09}. 
We use the ANN photo-z method, which has proved to 
be a superior training set method \citep{oya08}, and briefly compare the 
results using different photometric observables.
Since the SDSS photometric catalog covers a large area of sky, a number 
of deep spectroscopic galaxy samples with SDSS photometry are available 
to use as training sets, as shown in Fig.~\ref{dist.spec}.

\section{SDSS Photometric Catalog and Galaxy Selection}
\label{sel}

The SDSS comprises a large-area 
imaging survey of the north Galactic cap, a multi-epoch imaging survey of  
an equatorial stripe in the south Galactic cap, and a spectroscopic survey of 
roughly $10^6$ galaxies and $10^5$ quasars 
\citep{yor00}. 
The survey uses a dedicated, wide-field, 2.5m telescope \citep{gun98} at 
Apache Point Observatory, New Mexico. 
Imaging is carried out in drift-scan mode using a 142 mega-pixel camera 
\citep{gun06} that gathers data in five broad bands, $u g r i z$, spanning 
the range from 3,000 to 10,000 \AA \, \citep{fuk96}, with an effective exposure 
time of 54.1 seconds per band. 
The images are processed using specialized 
software \citep{lup01,sto02} and are 
astrometrically \citep{pie03} and photometrically \citep{hog01,tuc06} 
calibrated using observations of a set of primary standard stars 
\citep{smi02} observed on a neighboring 20-inch telescope.

The seventh SDSS Data Release (DR7) imaging footprint increased $\sim$ 22\%
when compared to the previous data release (DR6) which covers an essentially 
contiguous region of the north Galactic cap. The additional coverage includes
the small missing patches in the contiguous region of the north galactic cap,
and the stripes which are part of the SEGUE (Sloan Extension for Galactic 
Understanding and Exploration) survey. In any region where imaging runs overlap, one run is 
declared primary\footnote{For the precise definition of primary objects see 
{\tt http://cas.sdss.org/dr7/en/help/docs/glossary.asp\#P}} 
and is used for spectroscopic target selection; 
other runs are declared secondary. 
The area covered by the DR7 primary imaging survey, including the 
southern stripes, is $11,663 \textrm{ deg}^2$ \citep{aba09}.

The SDSS stripe along the celestial equator in the Southern Galactic Cap
(``Stripe 82'') was imaged multiple times in the Fall months. This was first carried out
to allow a co-addition of the the repeat imaging scans in order to reach
fainter magnitudes, roughly 2 mag fainter than the single SDSS scans (see Table~\ref{propphot}). 
The co-addition includes a total of 122 runs, covering any
given piece of the $\sim$ 250 deg$^2$ area between 20 and 40 times. The co-addition
runs are designated 106 and 206 under the \verb1Stripe821 database in the
Catalog Archive Server (CAS) (see the SDSS CasJobs website{\tt http://casjobs.sdss.org/casjobs/}). 
The reader can find a detailed description of the co-addition
in \citet{annis11}.

\begin{figure}
  \begin{minipage}[t]{85mm}
    \begin{center}
      \resizebox{85mm}{!}{\includegraphics[angle=0]{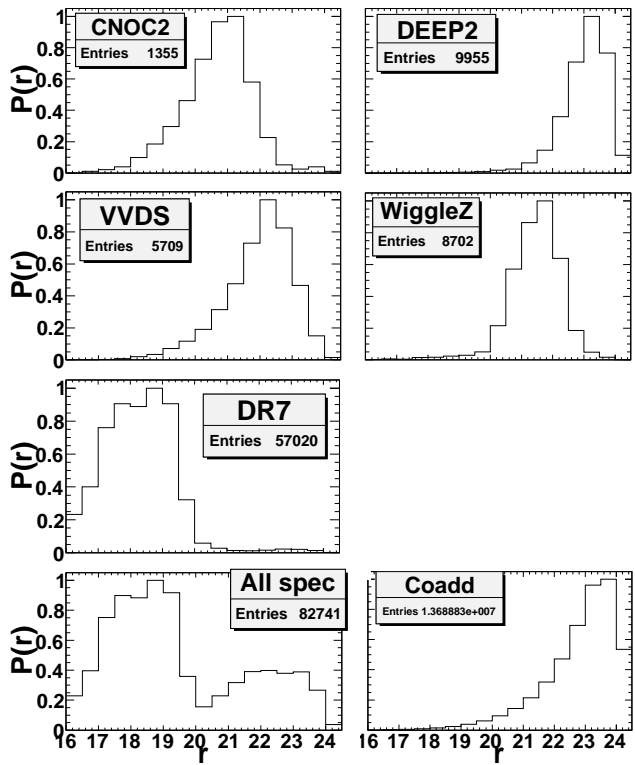}}
    \end{center}
  \end{minipage}
\caption{Normalized $r$ magnitude distributions for various catalogs.
  {\it Top three rows:} 
  the distributions of the spectroscopic catalogs used for photo-z
  training and validation are 
  shown for CNOC2, 
  DEEP2, VVDS, WiggleZ and SDSS DR7.  
  $Entries$ denotes the number of unique galaxy measurements used 
  from each catalog.
  {\it Bottom left:}
  distribution for the whole spectroscopic sample.
  {\it Bottom right:} the
  distribution for the SDSS coadd galaxy sample, where 
  objects were classified as galaxies according to the 
  photometric TYPE flag (see text).
}\label{dist.spec}
\end{figure}

The SDSS database provides a variety of measured magnitudes for each 
detected object. Throughout this paper, we use dereddened model magnitudes to 
perform the photometric redshift computations. To determine the model 
magnitude, the SDSS photometric pipeline fits two 
models to the image of each galaxy in each passband: a de Vaucouleurs (early-type) and 
an exponential (late-type) light profile. 
The models are convolved with the estimated point 
spread function (PSF), with arbitrary axis ratio and position angle.
The best-fit model in the $r$ band (which is used to fix the model scale 
radius) is then applied to the other passbands and convolved with the 
passband-dependent PSFs to yield the model magnitudes.
Model magnitudes provide an unbiased color estimate in the absence of color 
gradients \citep{sto02}, and the dereddening procedure removes the 
effect of Galactic extinction \citep{sch98}.

\begin{deluxetable}{c c}
\tablewidth{0pt}
\tablecaption{SDSS Coadd Properties}
\startdata
\hline
\hline
\multicolumn{2}{c}{\hspace{0.1 in} AB magnitude limits \hspace{0.2 in}  } \\
\hline
  \hspace{0.1 in}  $u$ & 23.25 \\ 
  \hspace{0.1 in}  $g$ & 23.51 \\ 
  \hspace{0.1 in}  $r$ & 23.26 \\ 
  \hspace{0.1 in}  $i$ & 22.69 \\ 
  \hspace{0.1 in}  $z$ & 21.27 \\ 
\enddata
\tablecomments{Magnitude limits are for 50\% completeness for galaxies
 in typical seeing \citep{annis11}. The median seeing for the SDSS imaging 
  survey is $1.4''$. 
} \label{propphot}
\end{deluxetable}

To construct the photometric sample of galaxies for which we wish to 
estimate photo-z's, we obtained
a catalog drawn from the SDSS CasJobs website.
We checked some of the SDSS photometric flags to ensure that we have obtained  
a reasonably clean galaxy sample. In particular, 
we selected all primary objects from \verb1Stripe821 that have the TYPE flag 
equal to $3$ (the type for galaxy) and that do not 
have any of the flags BRIGHT, SATURATED, or SATUR\_CENTER set.
For the definitions of these flags we refer the reader to the 
PHOTO flags entry at the SDSS 
website\footnote{{\tt http://cas.sdss.org/dr7/en/help/browser/browser.asp}}
or to Appendix \ref{query}.
We also took into account the nominal SDSS coadd flux limit
by only selecting galaxies with dereddened model 
magnitude $r<24.5$. In addition, the co-addition does not propagate information
on saturated pixels in individuals runs, and therefore the photometry
of objects brighter than $r=15.5$ is suspect. To circumvent this
issue we selected only galaxies with $r>16$.
The full database query we used is given in Appendix \ref{query}.

\begin{figure}
  \begin{minipage}[t]{85mm}
    \begin{center}
      \resizebox{85mm}{!}{\includegraphics[angle=0]{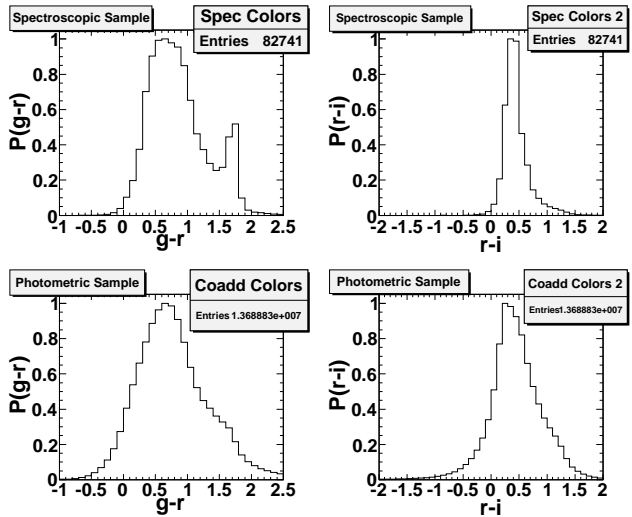}}
    \end{center}
  \end{minipage}
  \caption{Normalized distribution of $g-r$ and $r-i$ colors. 
  {\it Top row:} the color distributions for galaxies in the full spectroscopic 
    sample. 
    {\it Bottom row:} the color distributions for galaxies in the photometric 
    sample.
    As above, galaxy classification used the photometric TYPE flag.
}\label{dist.color.sdss}
\end{figure}

The final photometric sample comprises $13,688,828$ galaxies.
Only $2,267$ objects are in DR6 photometric redshift catalog
from \citet{oya08}.
The $r$ magnitude distribution of this sample is shown in 
the bottom right panel of Fig.~\ref{dist.spec}; the $g-r$ and 
$r-i$ color distributions 
are shown in the bottom panels of Fig.~\ref{dist.color.sdss}.

\section{Spectroscopic Training and Validation sets} \label{tra}

Since our methods to estimate photo-z's and photo-z errors are 
training-set based, we would ideally like the spectroscopic 
training set to be
fully representative of the photometric sample to be analyzed, i.e., to have
similar statistical properties and magnitude/redshift distributions.  
Training-set methods can be thought of as inherently Bayesian, in the sense 
that the training-set distributions form effective priors for the analysis of the 
photometric sample; to the extent that the training-set distributions 
reflect those of the photometric sample, we may expect the photo-z estimates 
to be unbiased (or at least they will not be biased by the prior). 
Given the practical difficulties of carrying out spectroscopy at 
faint magnitudes and low surface brightness, such an ideal generally cannot be achieved.
Realistically, all we can hope for is a training set that 
(a) is large enough that statistical fluctuations are small and (b) 
spans the same magnitude, color, and redshift ranges as the photometric sample
 \citep{oya08}. 

We have constructed a spectroscopic sample consisting of $82,741$ 
galaxies that have SDSS coadd photometry measurements and that have 
spectroscopic redshifts measured by the SDSS or by 
other surveys, as described below. 
We imposed a magnitude limit of $16<r<24.5$ on the spectroscopic 
sample and applied 
additional cuts on the quality of the spectroscopic
redshifts reported by the different surveys.
Each survey providing spectroscopic redshifts defines a redshift 
quality indicator; we refer the reader to the respective publications listed 
below for their precise definitions.
For each survey, we chose a redshift quality cut roughly corresponding
to 90\% redshift confidence or greater. 
The SDSS spectroscopic sample 
provides $57,020$ 
redshifts with confidence level 
$z_{\rm conf} > 0.9$. The remaining redshifts are: $1,355$ 
from the Canadian Network for Observational Cosmology 
Field Galaxy Survey \citep[CNOC2;][]{yee00}, $9,955$ 
from the Deep Extragalactic Evolutionary Probe (DEEP2)  
\citep{wei05}\footnote{{\tt http://deep.berkeley.edu/DR2/ }}
with $z_{\rm quality} \geq 3$, $8,702$ 
from the WiggleZ Dark Energy Survey
\citep{dri10} with $QoP > 3$, $5,709$ 
from the VIsible imaging Multi-Object Spectrograph - 
Very Large Telescope Deep Survey 
(VVDS) \citep{gal08} with flag $3$ and $4$. 

The spectroscopic sample obtained by combining all these catalogs 
 was divided into two catalogs of the 
same size ($\sim 42,000$ objects each). 
One of these catalogs was taken to be 
the {\it training set} used by the photo-z and error estimators, and the other
was used as a {\it validation set} to carry out tests of photo-z
quality (see \S \ref{subsec:meth_photoz}). 

The $r$-magnitude distributions for each spectroscopic
sample are shown in Fig. \ref{dist.spec}, while Fig. 
\ref{dist.color.sdss} shows the color ($g-r$ and $r-i$) 
distributions for all objects in the final spectroscopic
sample. As for how representative the spectroscopic 
training and validation sample are for the full photometric 
sample, we checked that the color/magnitude space is fully 
covered by the spectroscopic sample up to redshift $0.75$ - 
$0.8$. Beyond this redshift range, the spectroscopic sample 
partially cover the color/magnitude space. Therefore, the 
reader need to be cautious when using photo-z's beyond this 
range.

\section{Methods}\label{met}

\subsection{ANN Photometric redshifts} 
\label{subsec:meth_photoz}

The ANN method that we use to estimate galaxy photo-z's is 
a general classification and interpolation tool used 
successfully in a variety of fields.
We use a particular type of ANN called a Feed Forward Multilayer 
Perceptron to map the relationship between photometric observables 
and redshifts, as implemented in \citet{oya08}.

In this work we use $X$:15:15:15:1 networks to estimate photo-z's, where $X$ is the
number of input photometric parameters per galaxy, following the notation of 
\citet{col04}.
The corresponding number of degrees of freedom (the number of weights) is
roughly 1,000, depending on the actual value of $X$.

Following \citet{oya08}, in order to avoid over-fitting, the spectroscopic sample is divided into two 
independent subsets, the  
{\it training} and {\it validation} sets, 
and the formal minimizations are done using the training set.
After each minimization step, the network is evaluated on the 
validation set, and 
the set of weights that performs best on the validation set 
is chosen as the final set. 
To reduce the chance of ending in a less-than-optimal local minimum,
we minimize five networks starting at different positions in the space of weights.
Among these, we choose the network that gives the lowest photo-z scatter 
in the validation set. 

We calculated photo-z's using galaxy magnitudes, colors, and the
concentration indices for all passbands.  
The concentration index $c_i$ in a passband $i$ is defined as the ratio of {\tt PetroR50} 
and {\tt PetroR90}, which are the radii that encircle 50\% and 90\% of the 
Petrosian flux, respectively. Early-type (E and S0) galaxies, with centrally 
peaked surface brightness profiles, tend to have low values of the 
concentration index, while late-type spirals, with quasi-exponential light 
profiles, typically have higher values of $c$.
Previous studies \citep{morg58,shi01,yam05,par05} have shown 
that the concentration parameter correlates well
with galaxy morphological type, and we used it to help break the 
degeneracy between redshift and galaxy type.
We present the photo-z results for different combinations of input 
parameters in \S\ref{res}.

\subsection{Photometric redshift errors}\label{meter}

We estimated photo-z errors for objects in the photometric catalog using 
the Nearest Neighbor Error (NNE) estimator \citep{oya08b},
publicly available.\footnote{{\tt 
http://kobayashi.physics.lsa.umich.edu/$\sim$ccunha/nearest/}} 
The NNE method is training-set based, with
a neighbor selection similar to the NNP photo-z estimator; it 
associates photo-z errors to photometric objects by considering the 
errors for objects with similar multi-band magnitudes in the 
validation set. 
We use the validation set, because the photo-z's of the training set could be
over-fit, which would result in NNE underestimating the photo-z errors.
In studies of photo-z error estimators applied
to mock and real galaxy catalogs, \citet{oya08b} found that NNE  
accurately predicts the photo-z error when the training set is 
representative of the photometric sample. In the following, 
$\sigma^{NNE}_z$ will denote the nearest neighbors error
estimate.

\section{Results} \label{res}

To test the quality of the photo-z estimates, we use the photo-z
bias $z_{bias}$ and the photo-z RMS scatter, $\sigma$, defined
by
\begin{equation}
z_{bias}=\frac{1}{N}\sum^N_{i=1}(z_{phot,i}-z_{spec,i}),
\end{equation}
\begin{equation}
\sigma^2=\frac{1}{N}\sum^N_{i=1}(z_{phot,i}-z_{spec,i})^2,
\end{equation}
and $\sigma_{68}$, the range containing 68\% of the validation set 
objects in the distribution of $\delta z=z_{phot,i}-z_{spec,i}$.
In other words, $\sigma_{68}$ is the value of $|z_{phot,i}-z_{spec,i}|$
such that 68\% of the objects have $|z_{phot,i}-z_{spec,i}|<\sigma_{68}$.
Naturally, if the probability distribution function $P(\delta z)$
is Gaussian $\sigma$ and $\sigma_{68}$ coincide. We also consider
$\sigma_{95}$, defined in analogous way. 

We computed photo-z's using the ANN method with different combinations
of input photometric observables. All tested combinations are listed in
Table \ref{table:ANNcases}. In case M, we use the five magnitudes $ugriz$.
In case C, we use the four colors $u-g$, $g-r$, $r-i$ and $i-z$. In 
case CC, we use the four colors with the concentration indices 
$c_uc_gc_rc_ic_z$. We also repeat the cases M, C and CC splitting the
training set and the photometric sample into 4 bins of $r$ magnitude,
$r\leq 18$, $18< r\leq 20$, $20< r\leq 22$, $22< r\leq 24.5$, 
and perform separate ANN fits in each bin. These cases are dubbed Msplit,
Csplit and CCsplit, respectively. For all cases we use the same network
configuration, described in Section \ref{subsec:meth_photoz}.

\begin{figure*}
  \begin{center}
    \begin{minipage}[t]{50mm}
      \begin{center}
      \resizebox{50mm}{!}{\includegraphics[angle=0]{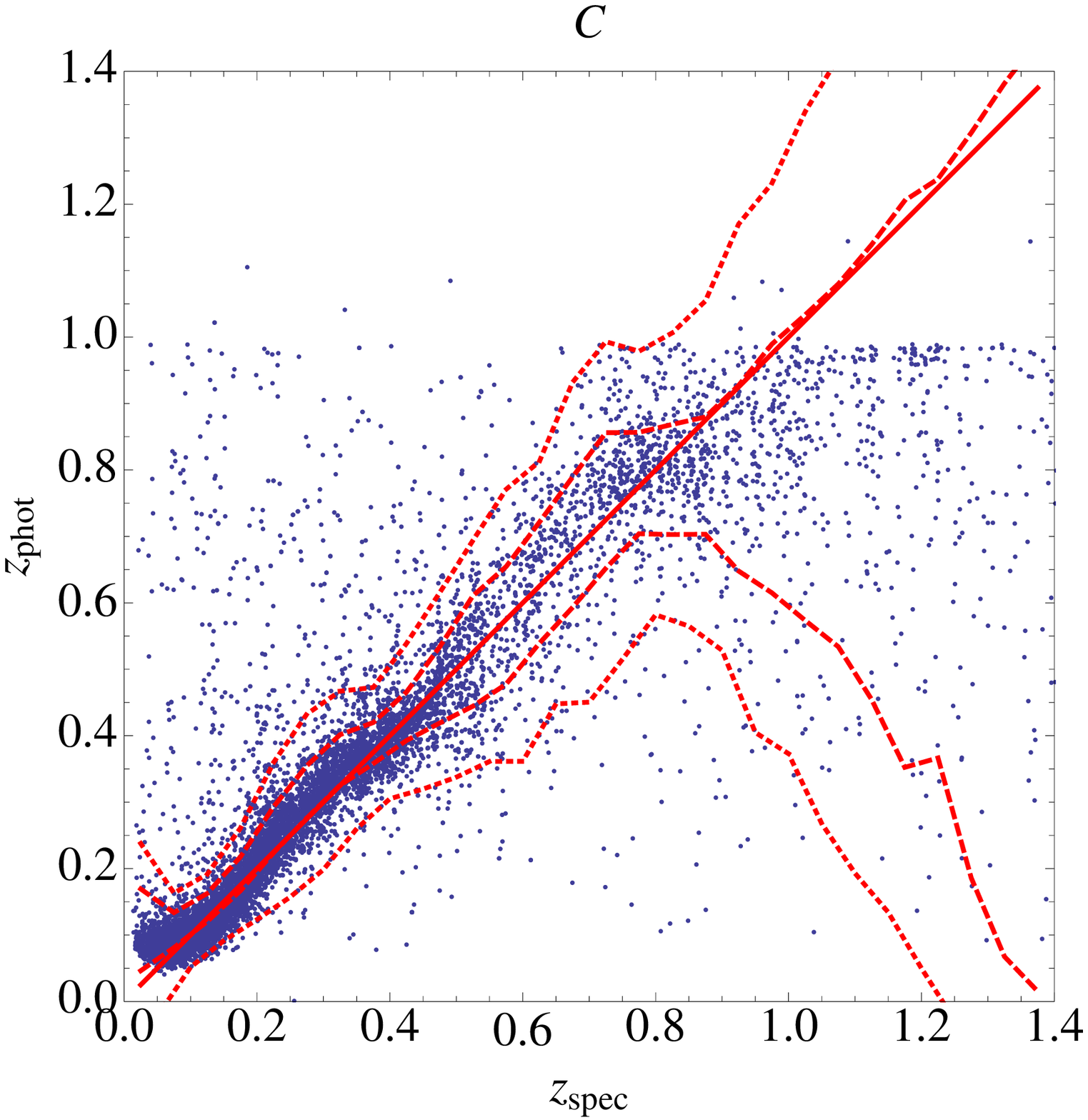}}
      \end{center}
    \end{minipage}
    \begin{minipage}[t]{50mm}
      \begin{center}
      \resizebox{50mm}{!}{\includegraphics[angle=0]{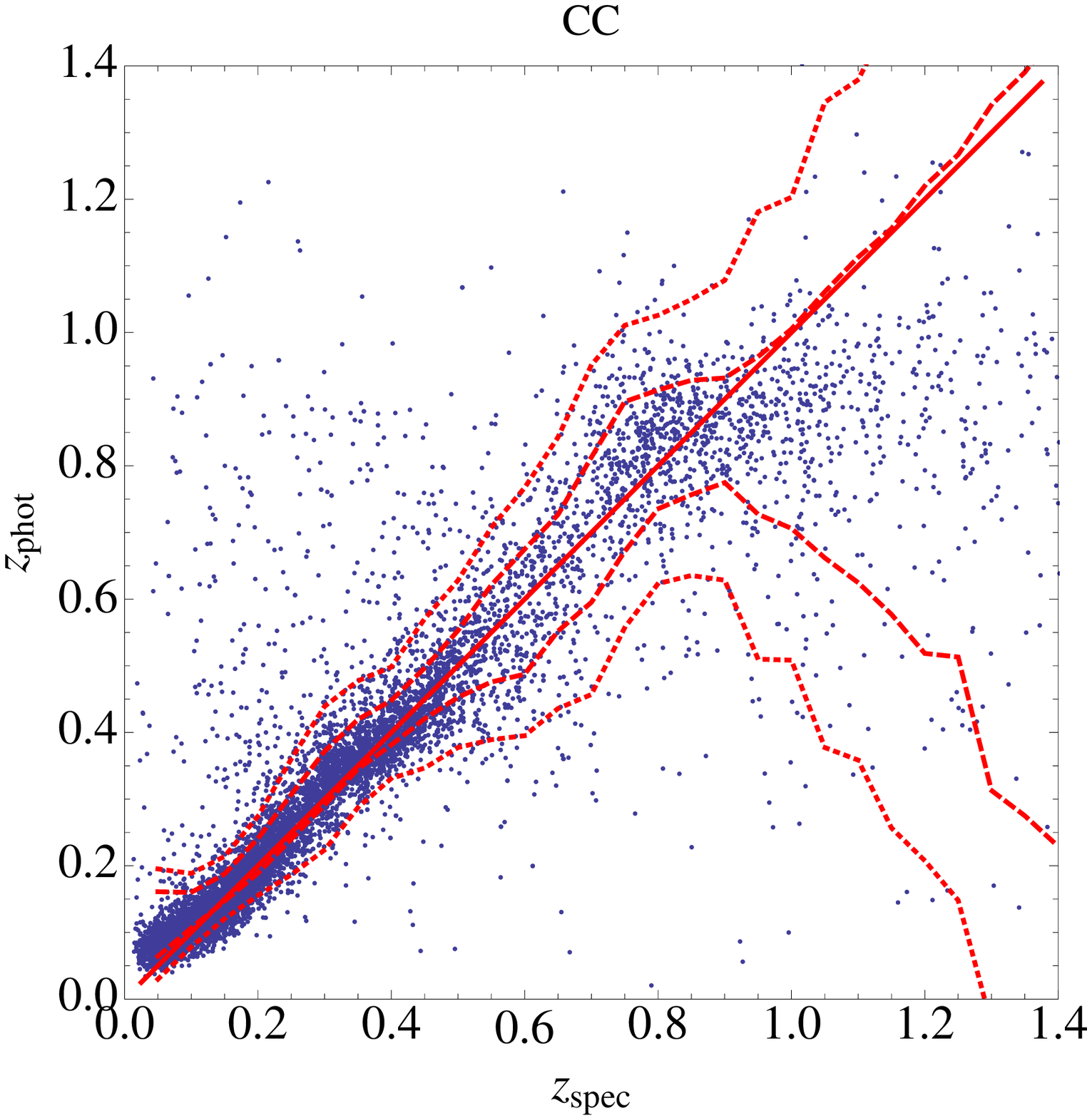}}
      \end{center}
    \end{minipage}
    \begin{minipage}[t]{50mm}
      \begin{center}
      \resizebox{50mm}{!}{\includegraphics[angle=0]{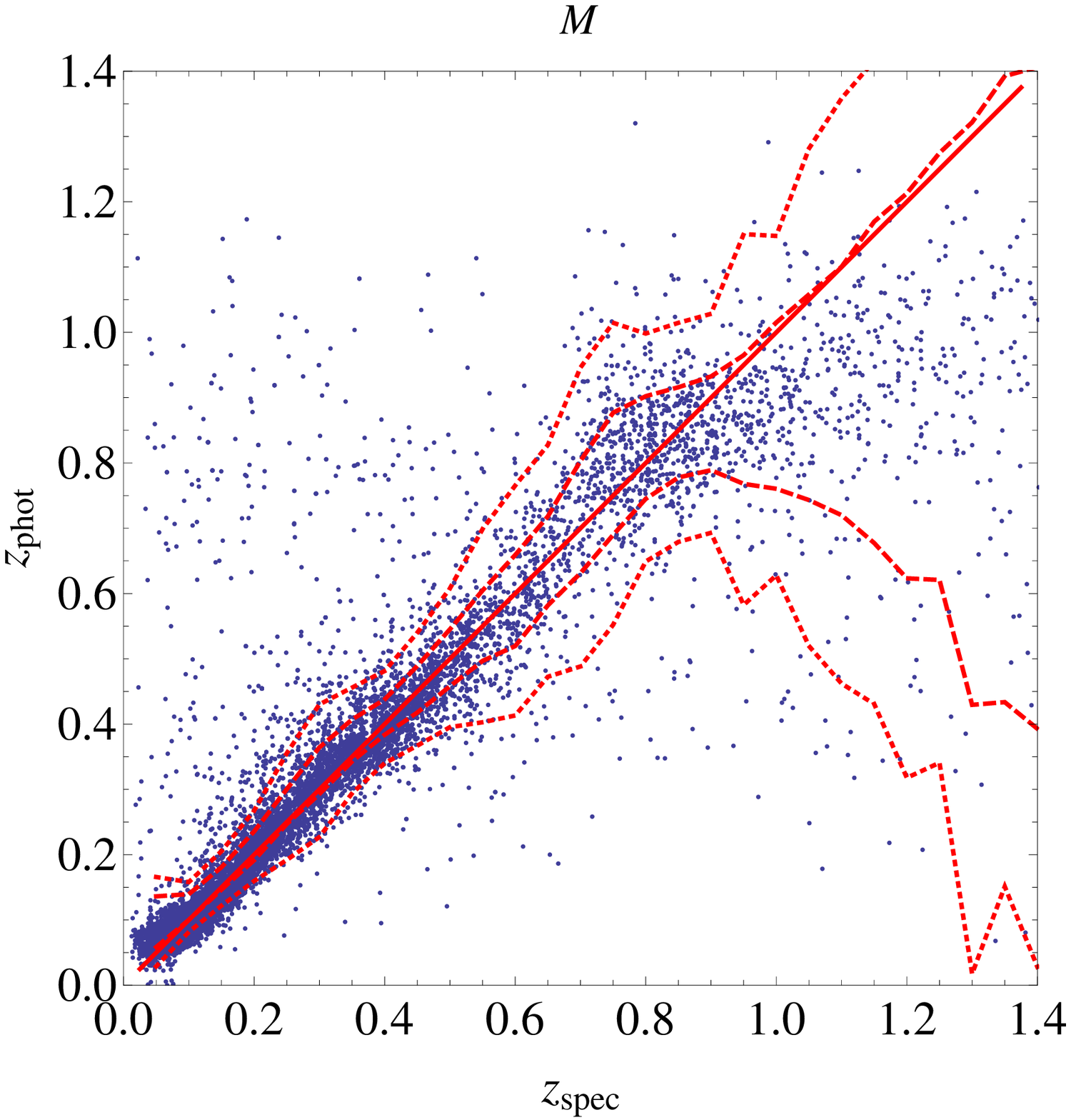}}
      \end{center}
    \end{minipage} \\
    \begin{minipage}[t]{50mm}
      \begin{center}
      \resizebox{50mm}{!}{\includegraphics[angle=0]{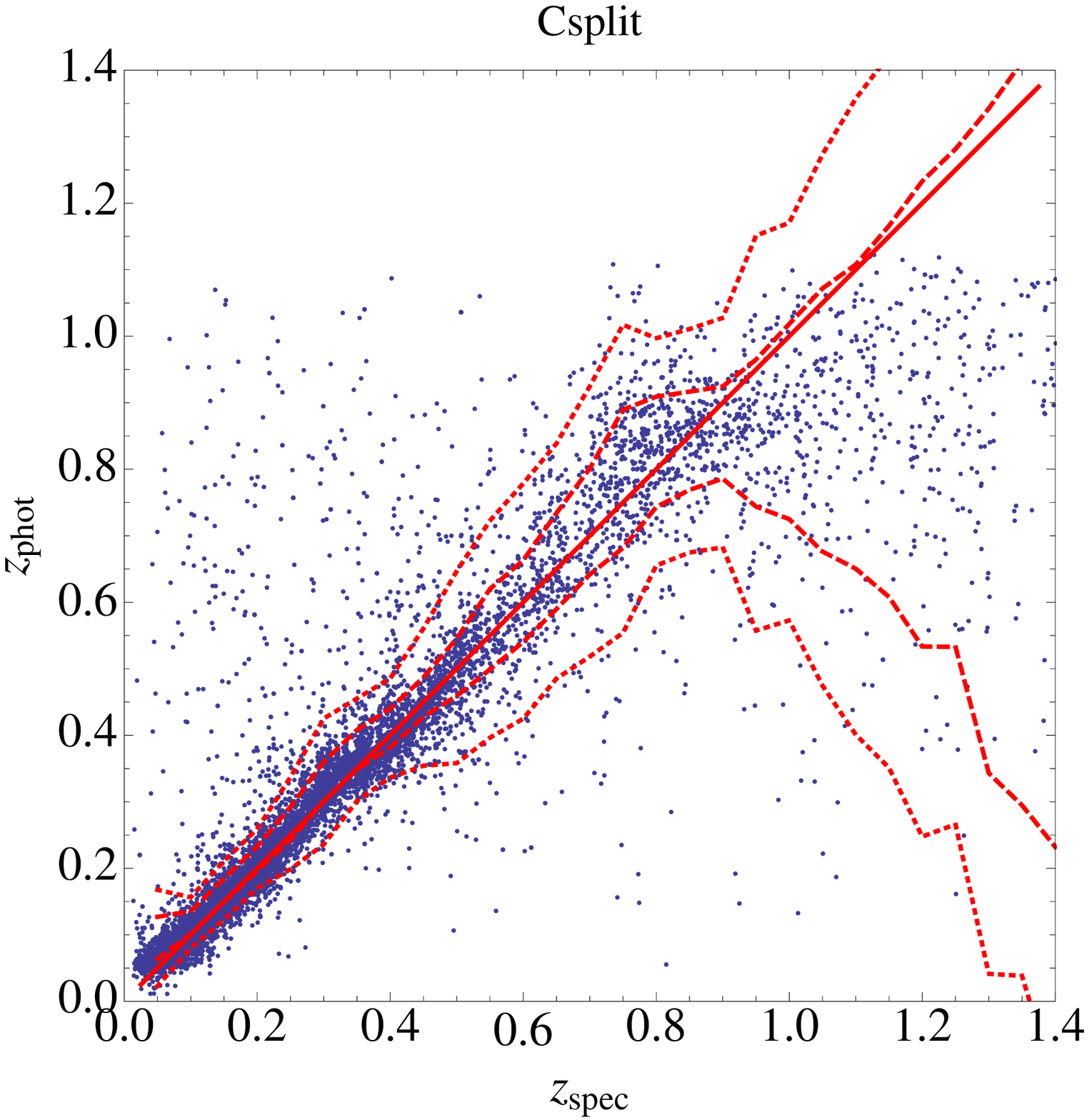}}
      \end{center}
    \end{minipage} 
    \begin{minipage}[t]{50mm}
      \begin{center}
      \resizebox{50mm}{!}{\includegraphics[angle=0]{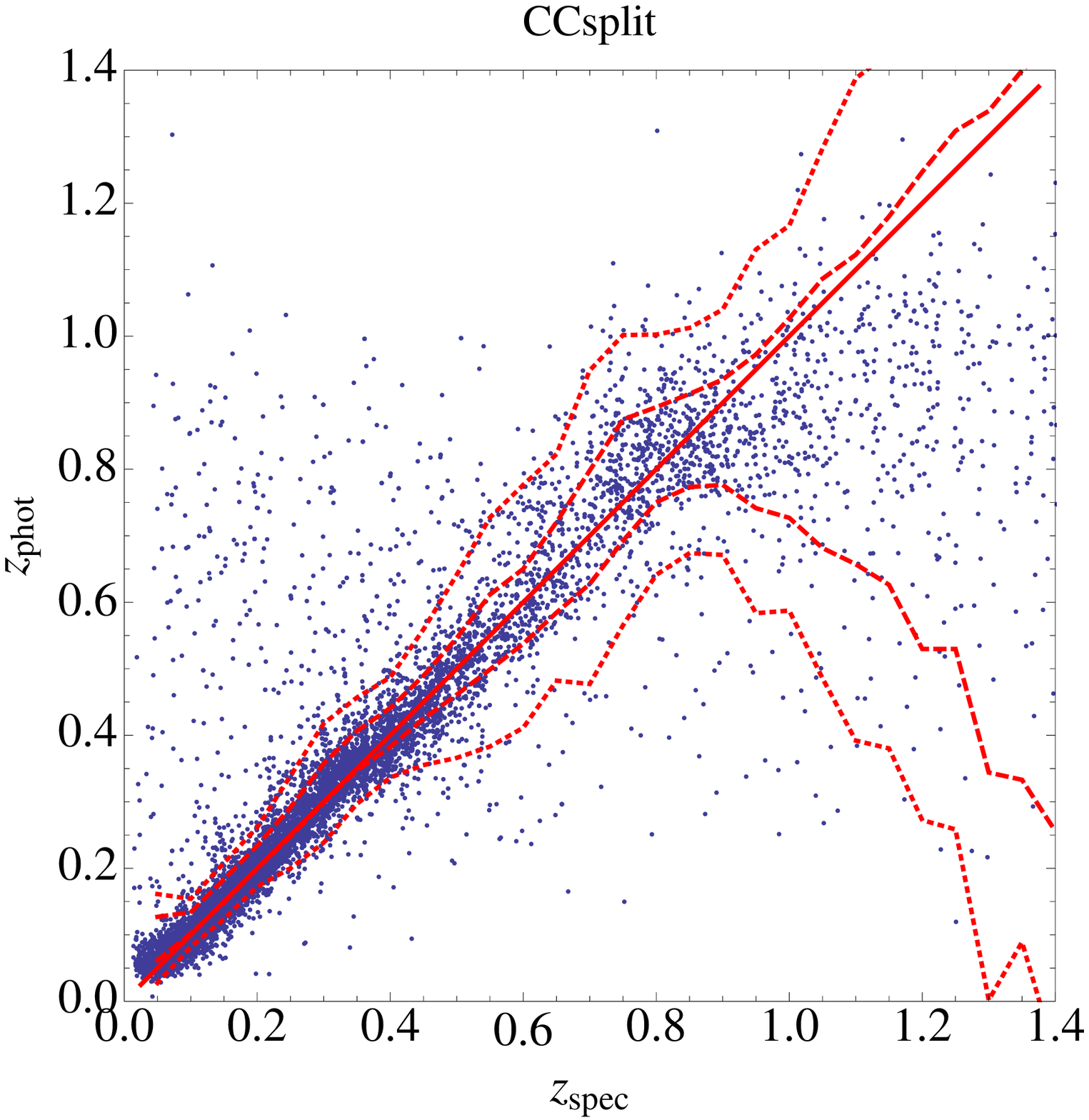}}
      \end{center}
    \end{minipage}
    \begin{minipage}[t]{50mm}
      \begin{center}
      \resizebox{50mm}{!}{\includegraphics[angle=0]{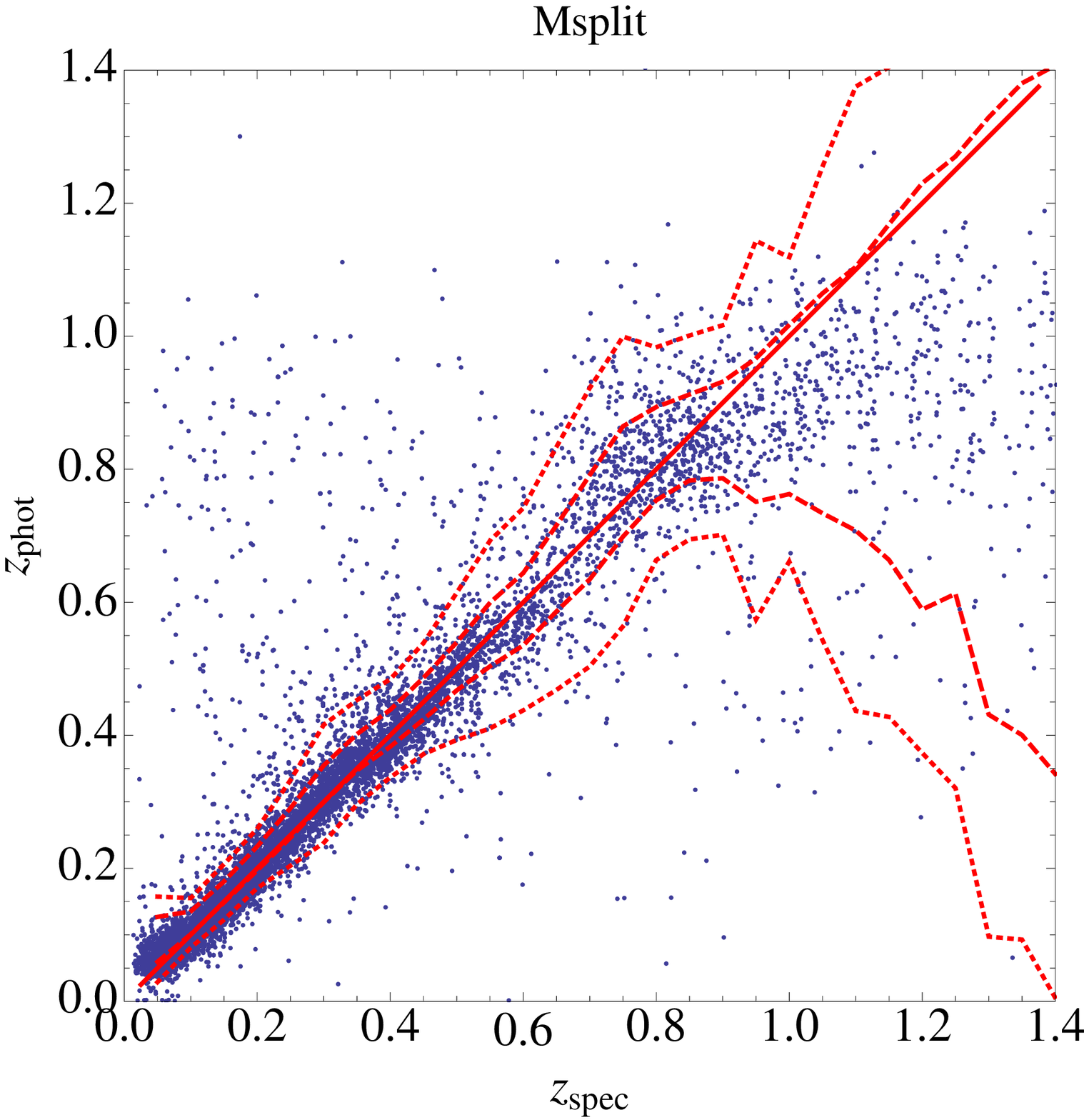}}
      \end{center}
    \end{minipage} 
  \end{center}
 \caption{ $z_{\rm phot}$ versus $z_{\rm spec}$ for the validation set for 
   different spectroscopic sets and different choices of photometric observables. 
   {\it Top Left:} Case C, where the input photometric data comprise 
   the 4 colors ($u-g$, $g-r$, $r-i$, $i-z$) 
   {\it Top Middle:} Case CC, where the input data are 
   the 4 colors $u-g$, $g-r$, $r-i$, $i-z$, and 5 concentration parameters $c_uc_gc_rc_ic_z$.
   {\it Top Right:} Case M, where we use only magnitudes.
   {\it Bottom Left:} Case Csplit, where we split the sample in $r$ magnitude slices.
   {\it Bottom Middle:} Case CCsplit, where we split the sample in $r$ magnitude slices.
   {\it Bottom Right:} Case Msplit, where we split the sample in $r$ magnitude slices.
   The solid line in each panel indicates $z_{\rm phot}=z_{\rm spec}$; the 
   dashed and dotted lines show the 68\% and 95\% confidence regions as a function
   of $z_{\rm spec}$ ($\sigma_{68}$ and $\sigma_{95}$), respectively.  
   The points display results for a random $10,000$ objects subset of the validation set. 
 }
\label{zpzs.valid_all}
\end{figure*}

In Fig. \ref{zpzs.valid_all} we plot the photometric redshift, $z_{phot}$,
for 10,000 randomly selected objects from validation set vs. true 
spectroscopic redshift, $z_{spec}$, for all considered cases.
In each panel, the solid line traces $z_{phot}=z_{spec}$ and the 
dashed and dotted lines show the corresponding 68\% and 95\% regions 
($\sigma_{68}$ and $\sigma_{95}$),
respectively, defined in $z_{spec}$ bins. We find that all cases produce very similar
results, in agreement with \citet{oya08}.

\begin{figure*}
  \begin{center}
    \begin{minipage}[t]{80mm}
      \begin{center}
      \resizebox{80mm}{!}{\includegraphics[angle=0]{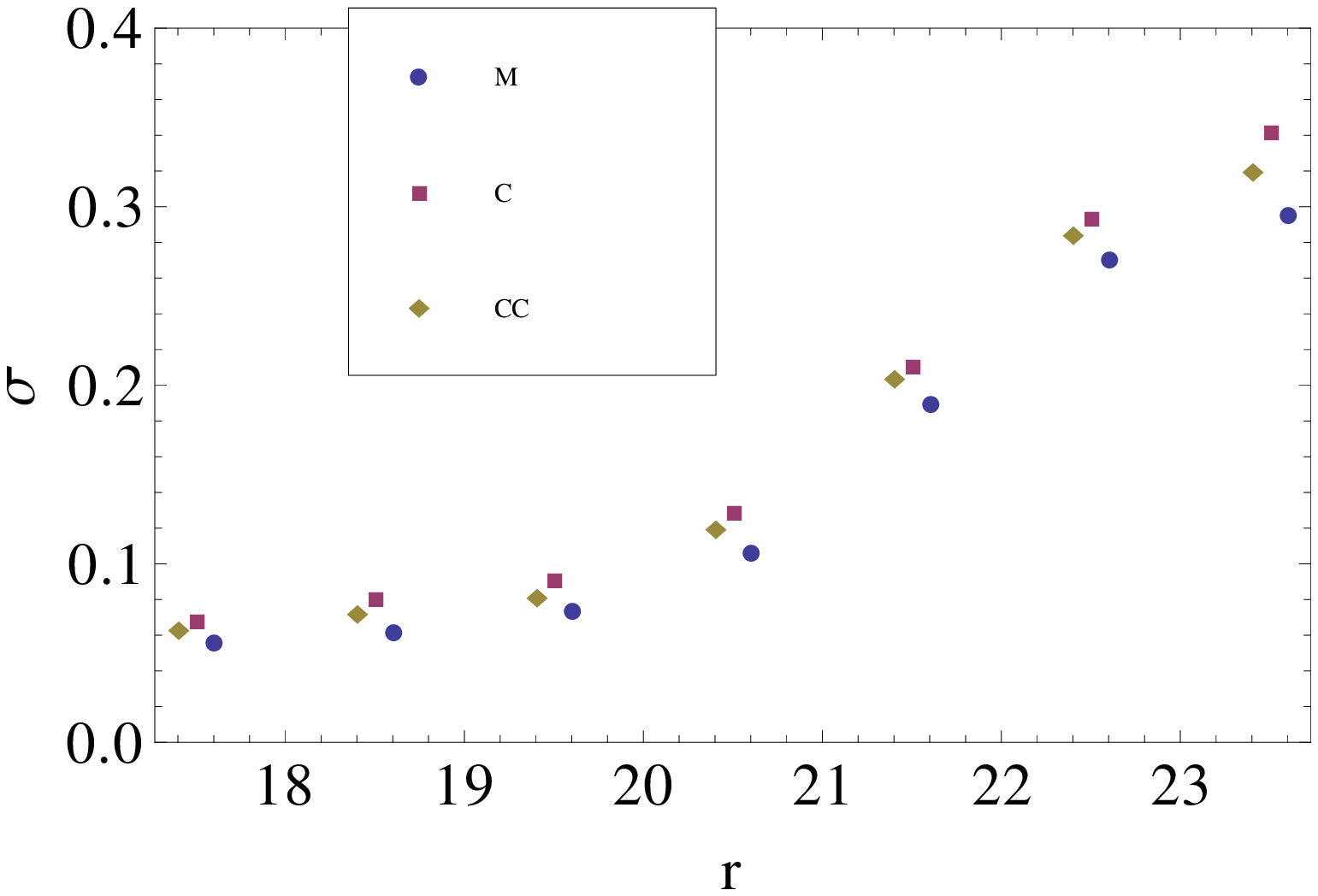}}
      \end{center}
    \end{minipage}
    \begin{minipage}[t]{80mm}
      \begin{center}
      \resizebox{80mm}{!}{\includegraphics[angle=0]{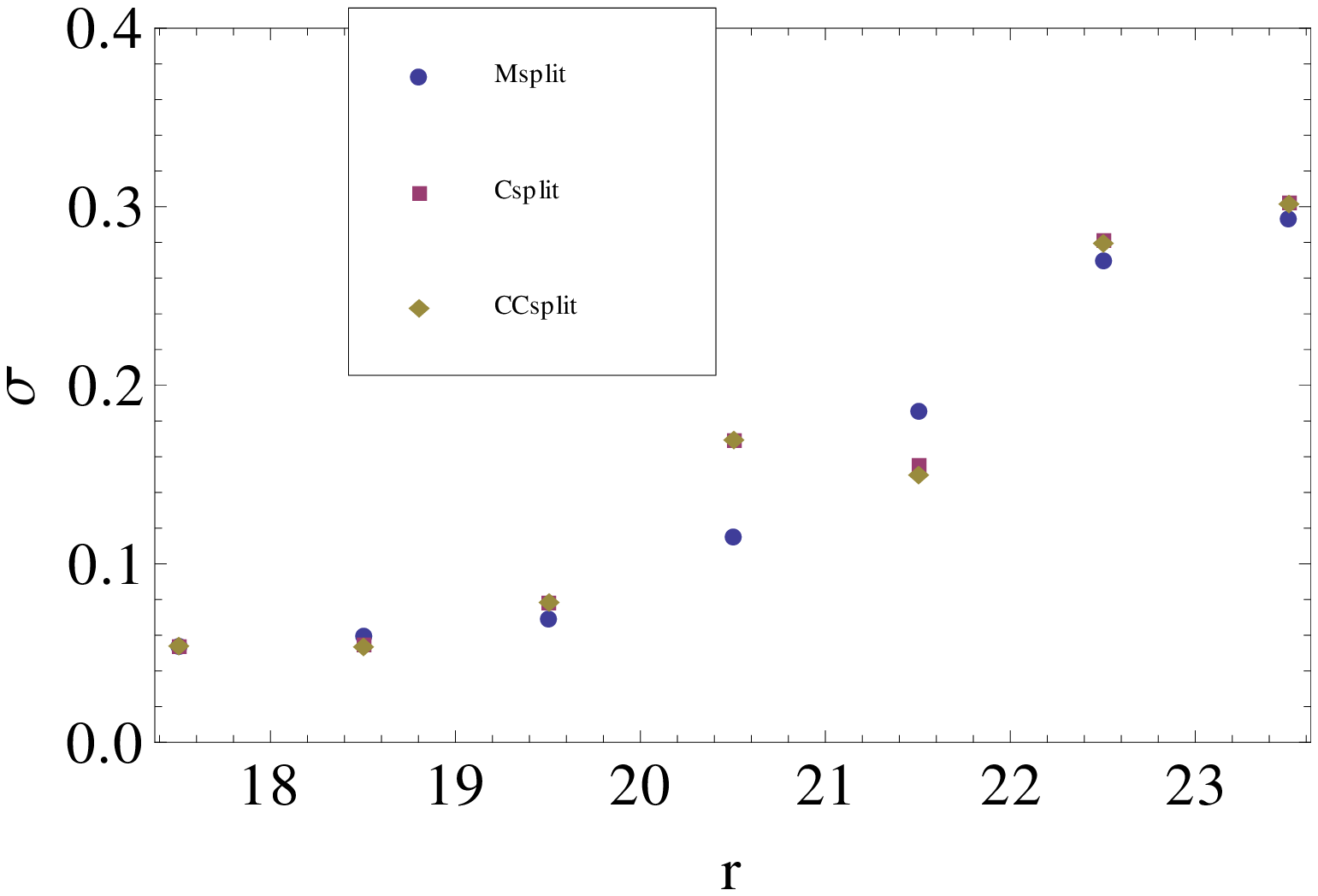}}
      \end{center}
    \end{minipage} \\
    \begin{minipage}[t]{80mm}
      \begin{center}
      \resizebox{80mm}{!}{\includegraphics[angle=0]{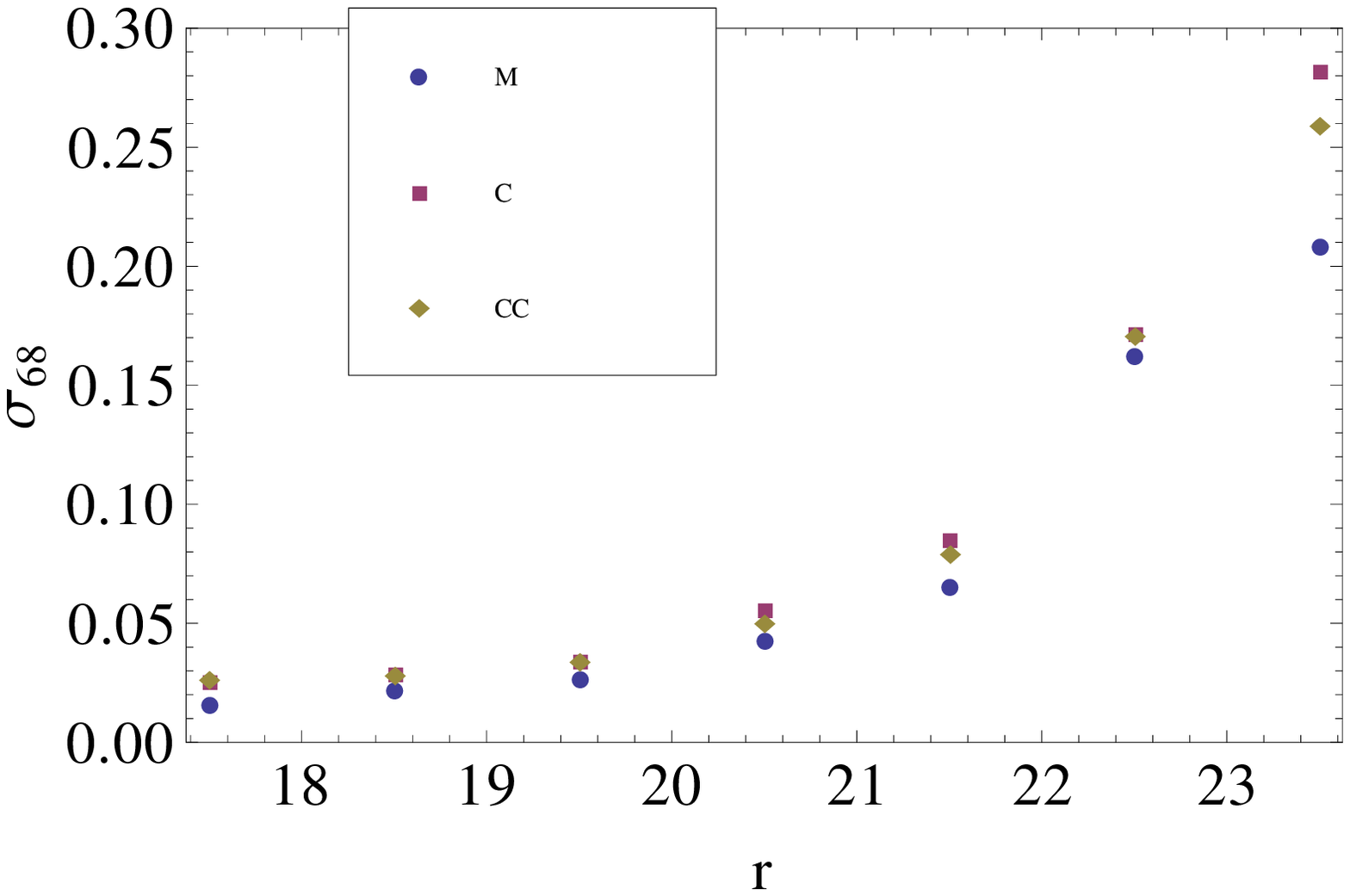}}
      \end{center}
    \end{minipage}
    \begin{minipage}[t]{80mm}
      \begin{center}
      \resizebox{80mm}{!}{\includegraphics[angle=0]{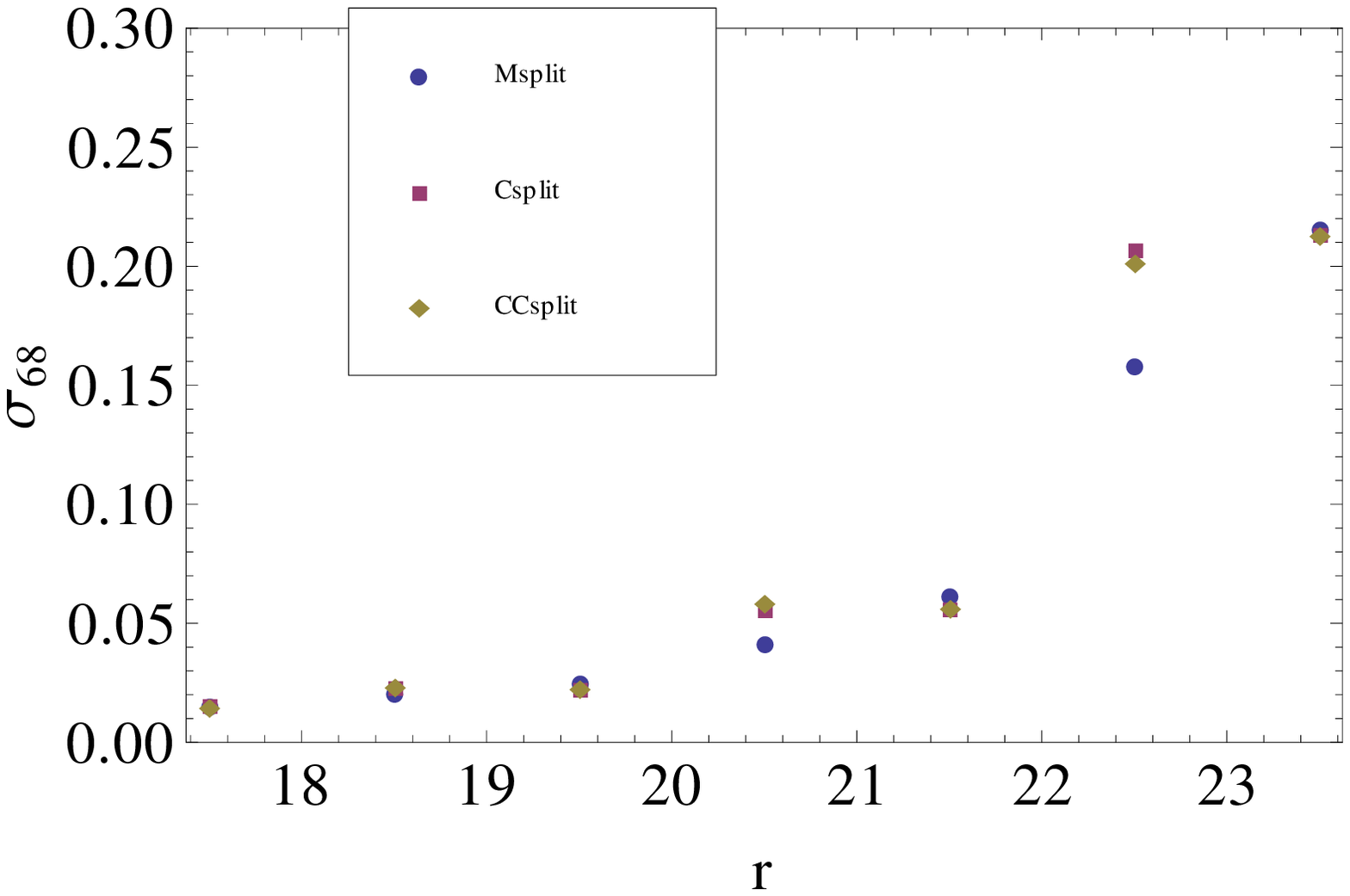}}
      \end{center}
    \end{minipage} 
  \end{center}
 \caption{$\sigma$ and $\sigma_{68}$ as functions of $r$ magnitude for all tested cases. 
 }
\label{sig:vs:mag}
\end{figure*}

\begin{figure*}
  \begin{center}
    \begin{minipage}[t]{80mm}
      \begin{center}
      \resizebox{80mm}{!}{\includegraphics[angle=0]{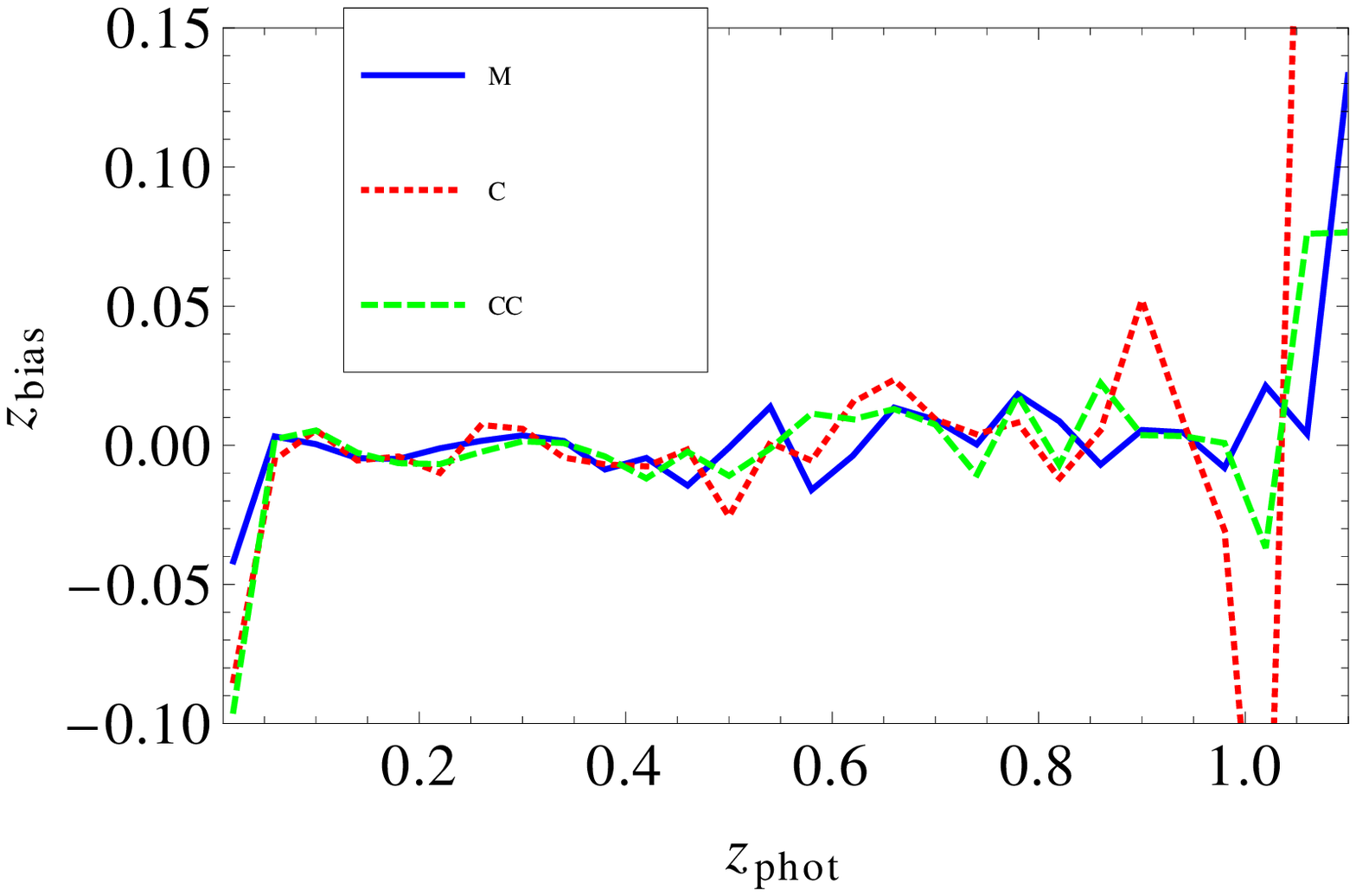}}
      \end{center}
    \end{minipage}
    \begin{minipage}[t]{80mm}
      \begin{center}
      \resizebox{80mm}{!}{\includegraphics[angle=0]{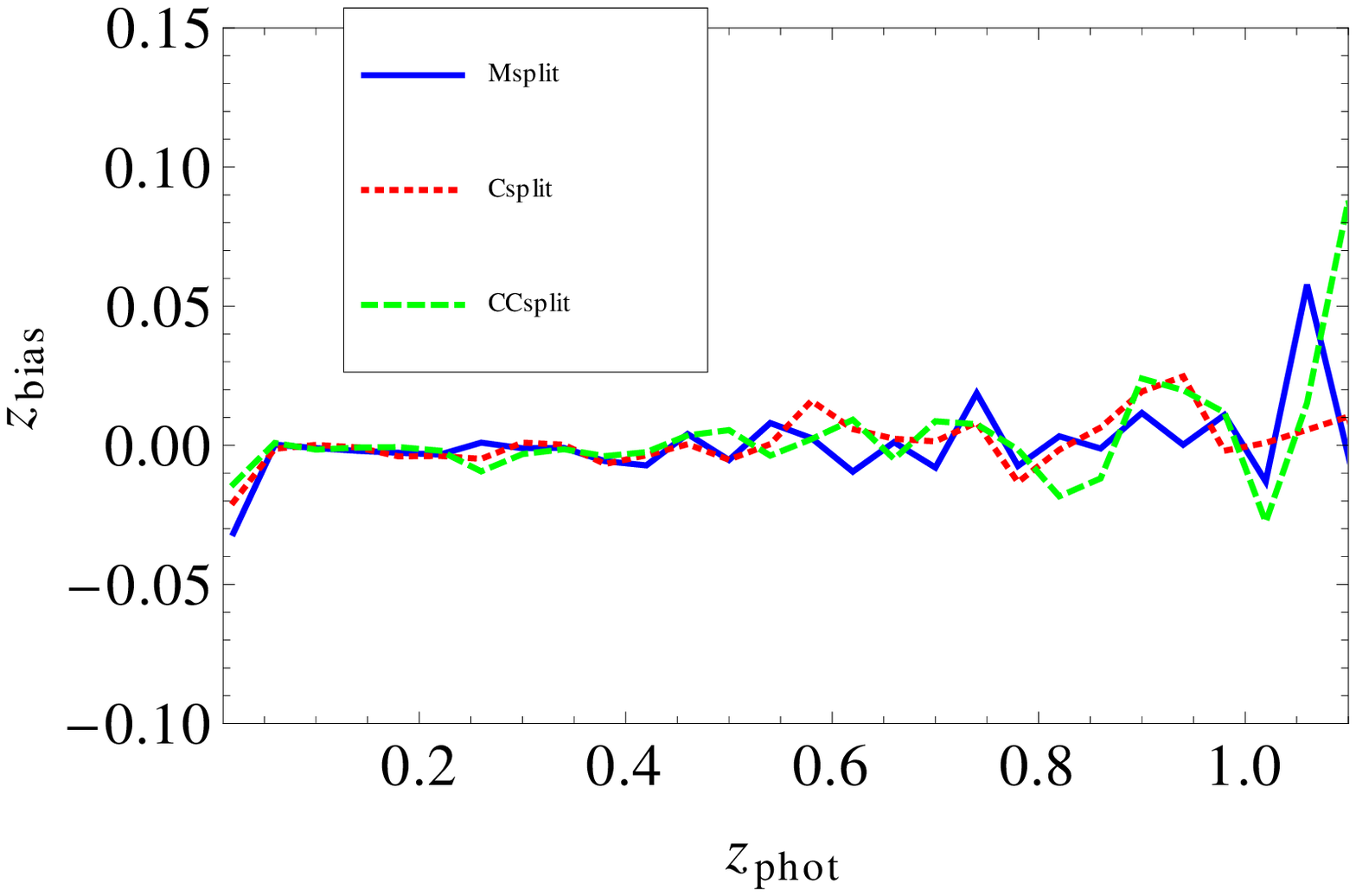}}
      \end{center}
    \end{minipage} 
  \end{center}
 \caption{$z_{bias}$ as a function of the photometric redshift for all tested cases. 
 }
\label{bias_vs_zphot}
\end{figure*}

\begin{figure*}
  \begin{center}
    \begin{minipage}[t]{80mm}
      \begin{center}
      \resizebox{80mm}{!}{\includegraphics[angle=0]{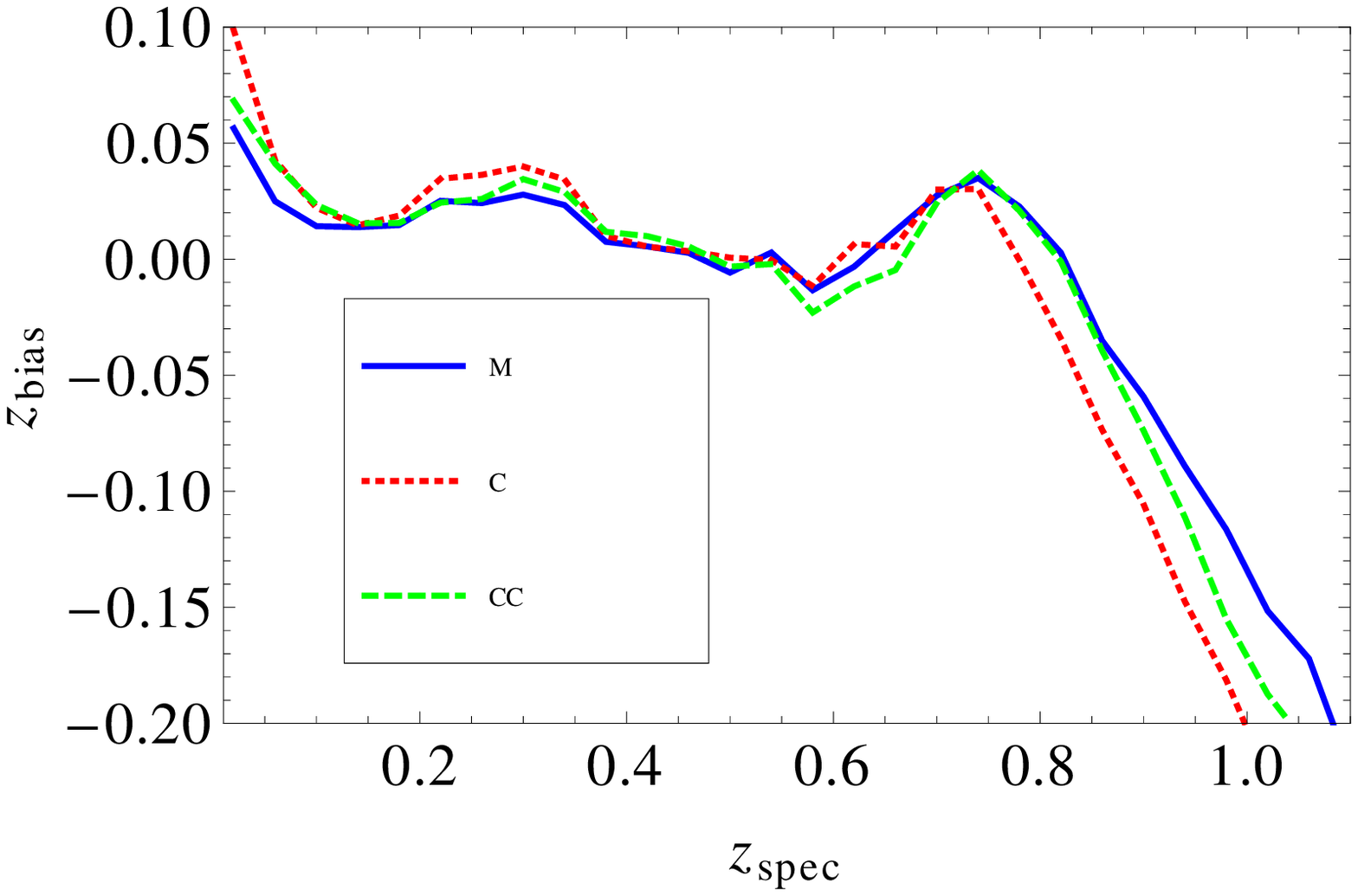}}
      \end{center}
    \end{minipage}
    \begin{minipage}[t]{80mm}
      \begin{center}
      \resizebox{80mm}{!}{\includegraphics[angle=0]{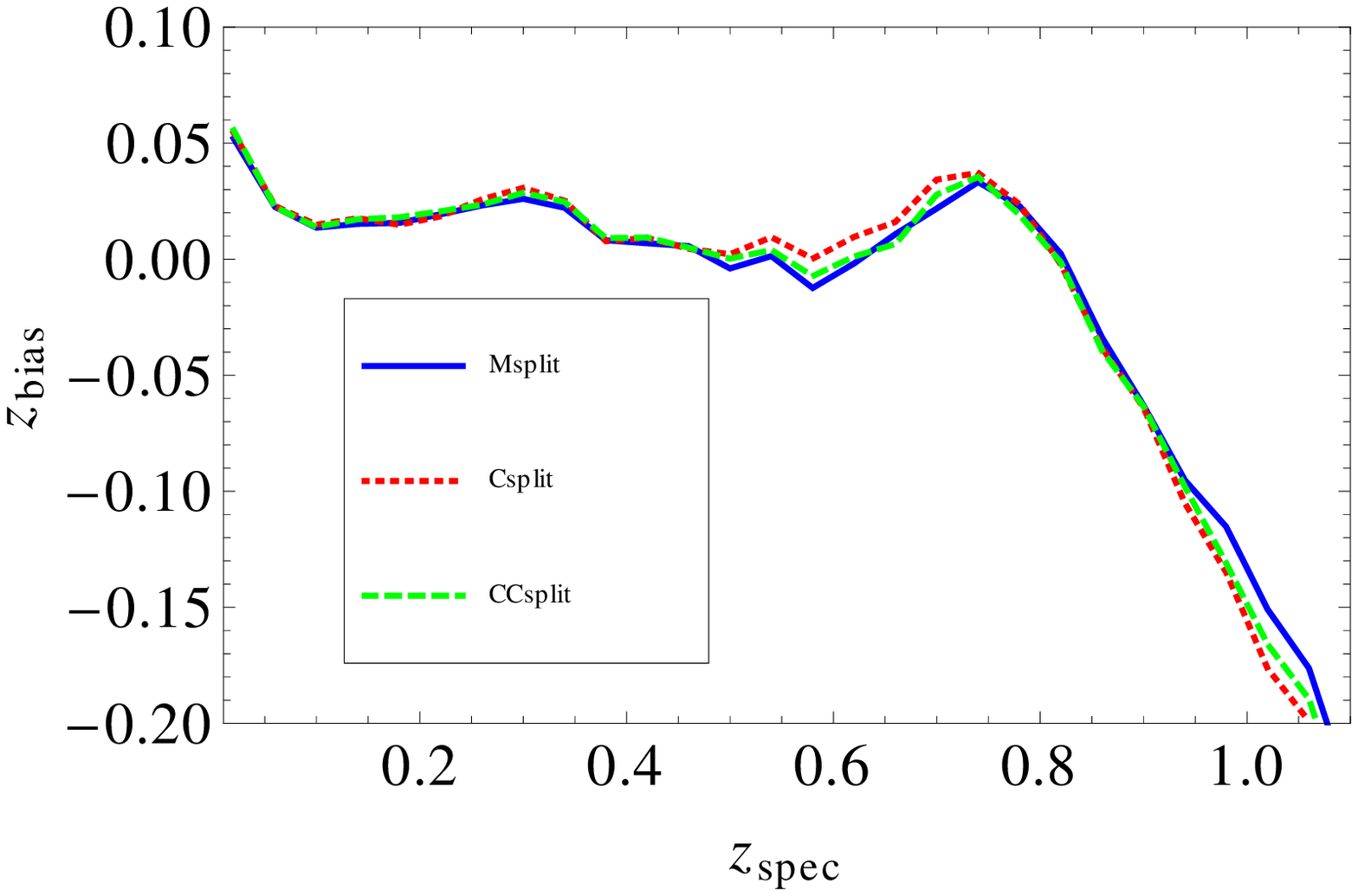}}
      \end{center}
    \end{minipage} 
  \end{center}
 \caption{$z_{bias}$ as a function of the spectroscopic redshift for all tested cases. 
 }
\label{bias_vs_zspec}
\end{figure*}

\begin{figure*}
  \begin{center}
    \begin{minipage}[t]{80mm}
      \begin{center}
      \resizebox{80mm}{!}{\includegraphics[angle=0]{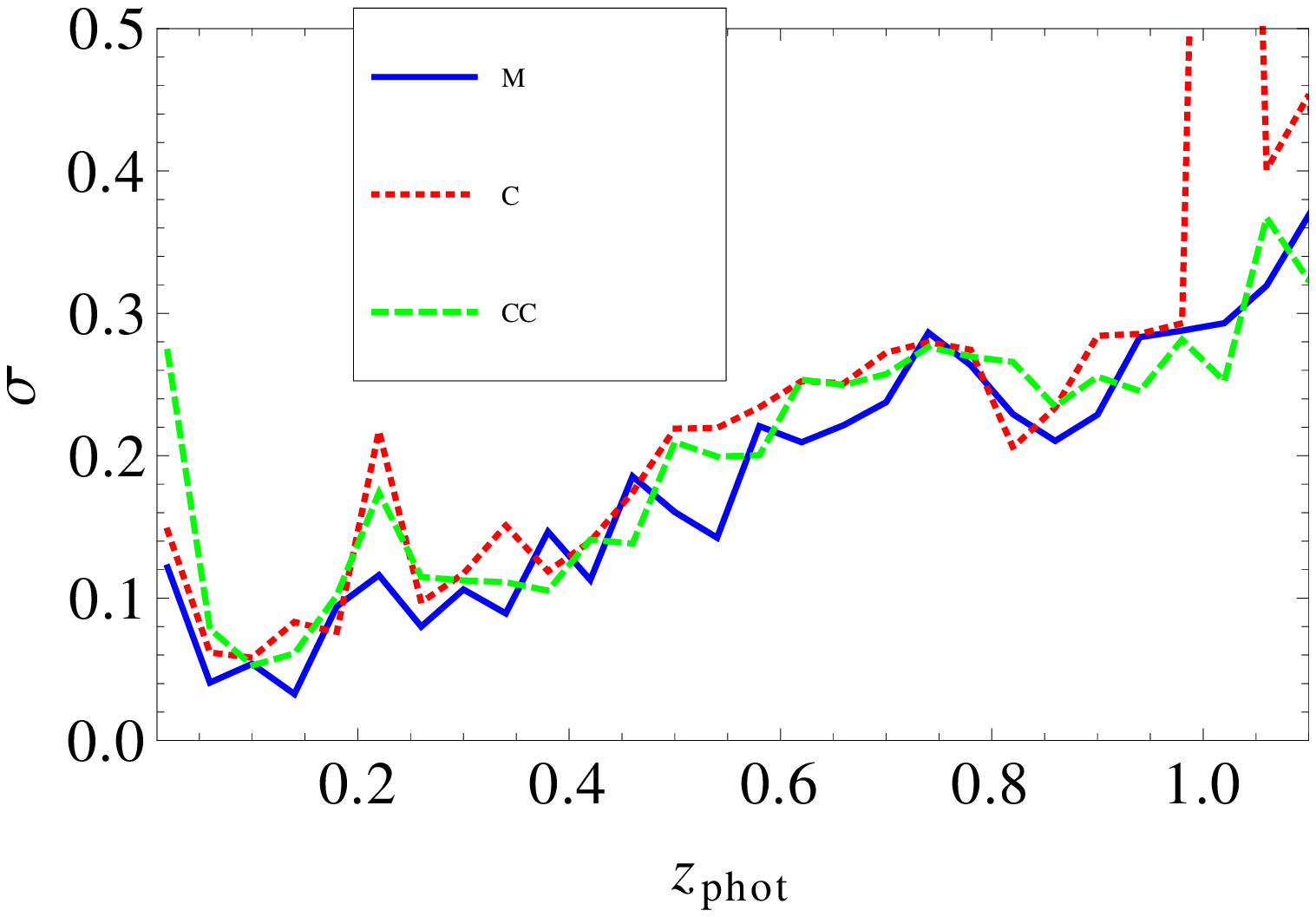}}
      \end{center}
    \end{minipage}
    \begin{minipage}[t]{80mm}
      \begin{center}
      \resizebox{80mm}{!}{\includegraphics[angle=0]{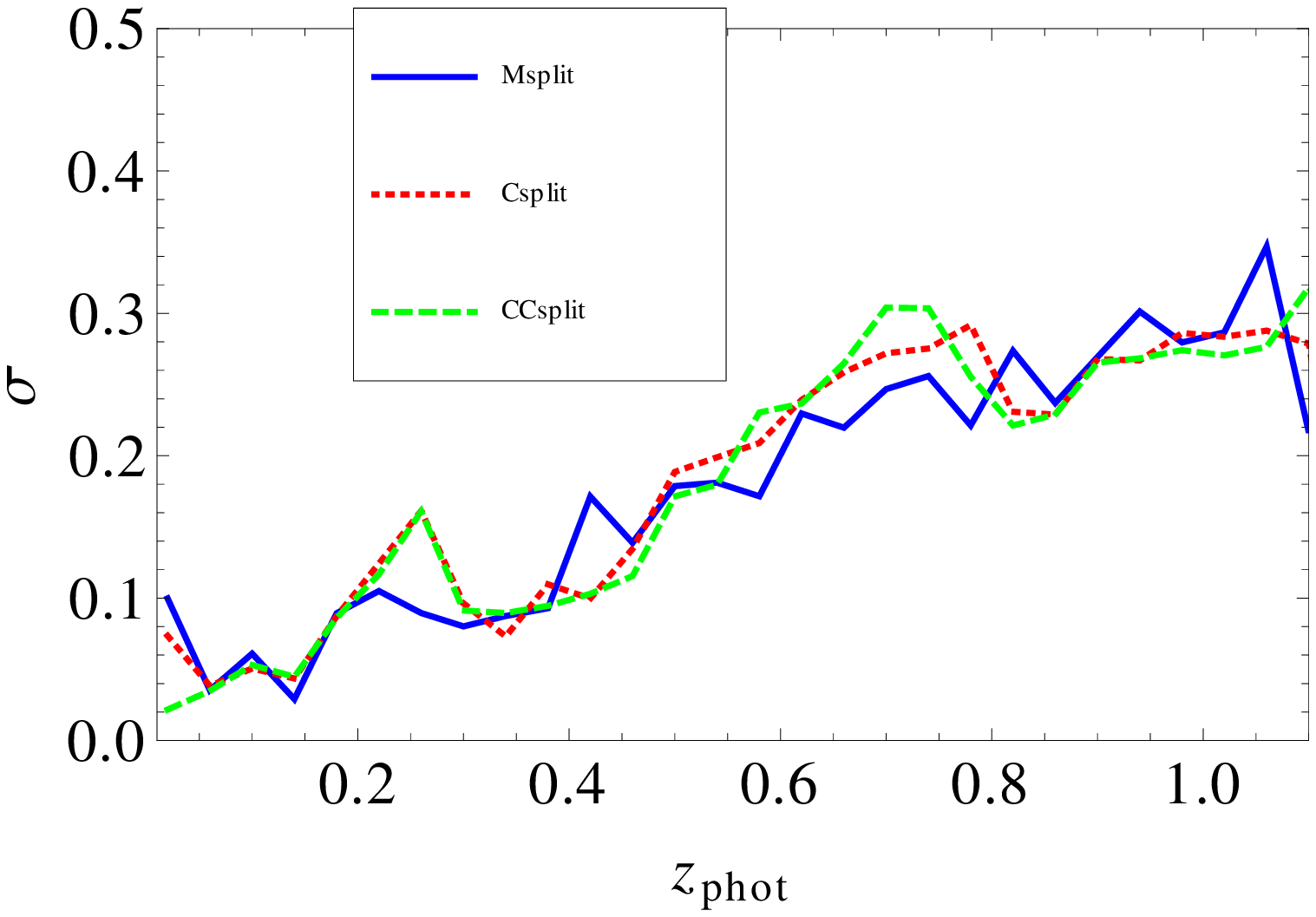}}
      \end{center}
    \end{minipage} \\
    \begin{minipage}[t]{80mm}
      \begin{center}
      \resizebox{80mm}{!}{\includegraphics[angle=0]{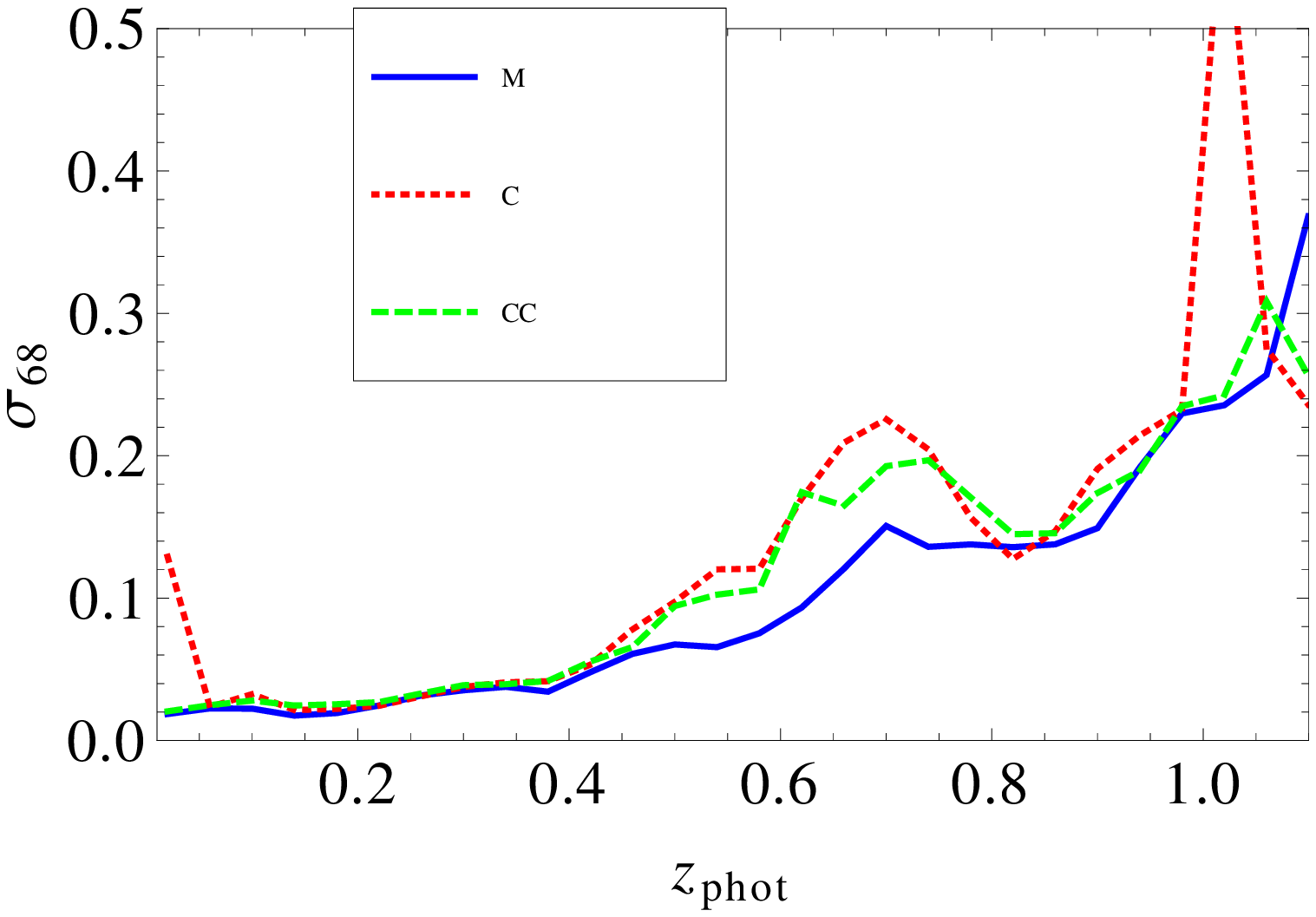}}
      \end{center}
    \end{minipage}
    \begin{minipage}[t]{80mm}
      \begin{center}
      \resizebox{80mm}{!}{\includegraphics[angle=0]{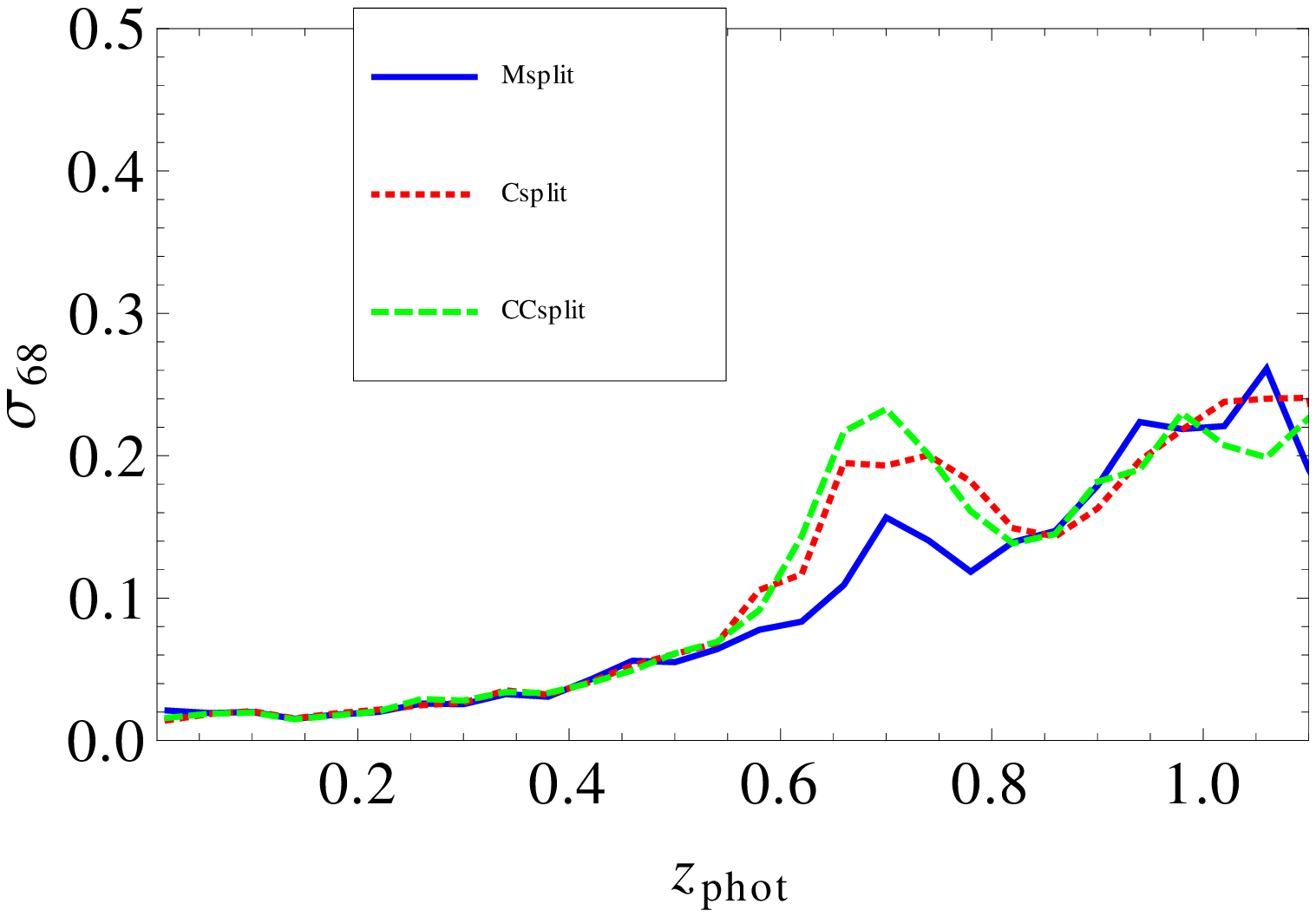}}
      \end{center}
    \end{minipage} 
  \end{center}
 \caption{$\sigma$ and $\sigma_{68}$ as a function of the photometric 
 redshift for all tested cases. 
 }
\label{sigma.vs.zphot}
\end{figure*}

\begin{figure*}
  \begin{center}
    \begin{minipage}[t]{80mm}
      \begin{center}
      \resizebox{80mm}{!}{\includegraphics[angle=0]{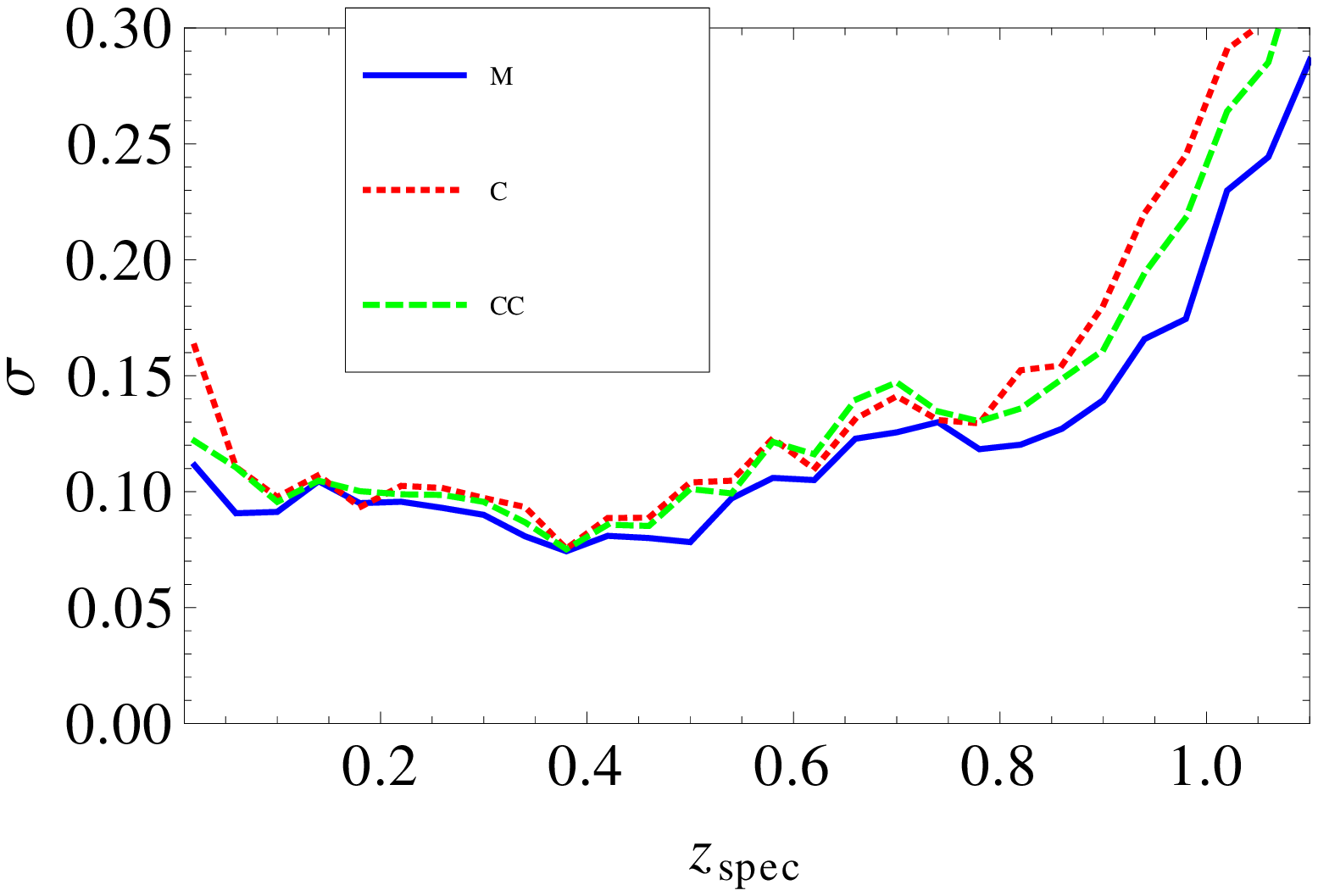}}
      \end{center}
    \end{minipage}
    \begin{minipage}[t]{80mm}
      \begin{center}
      \resizebox{80mm}{!}{\includegraphics[angle=0]{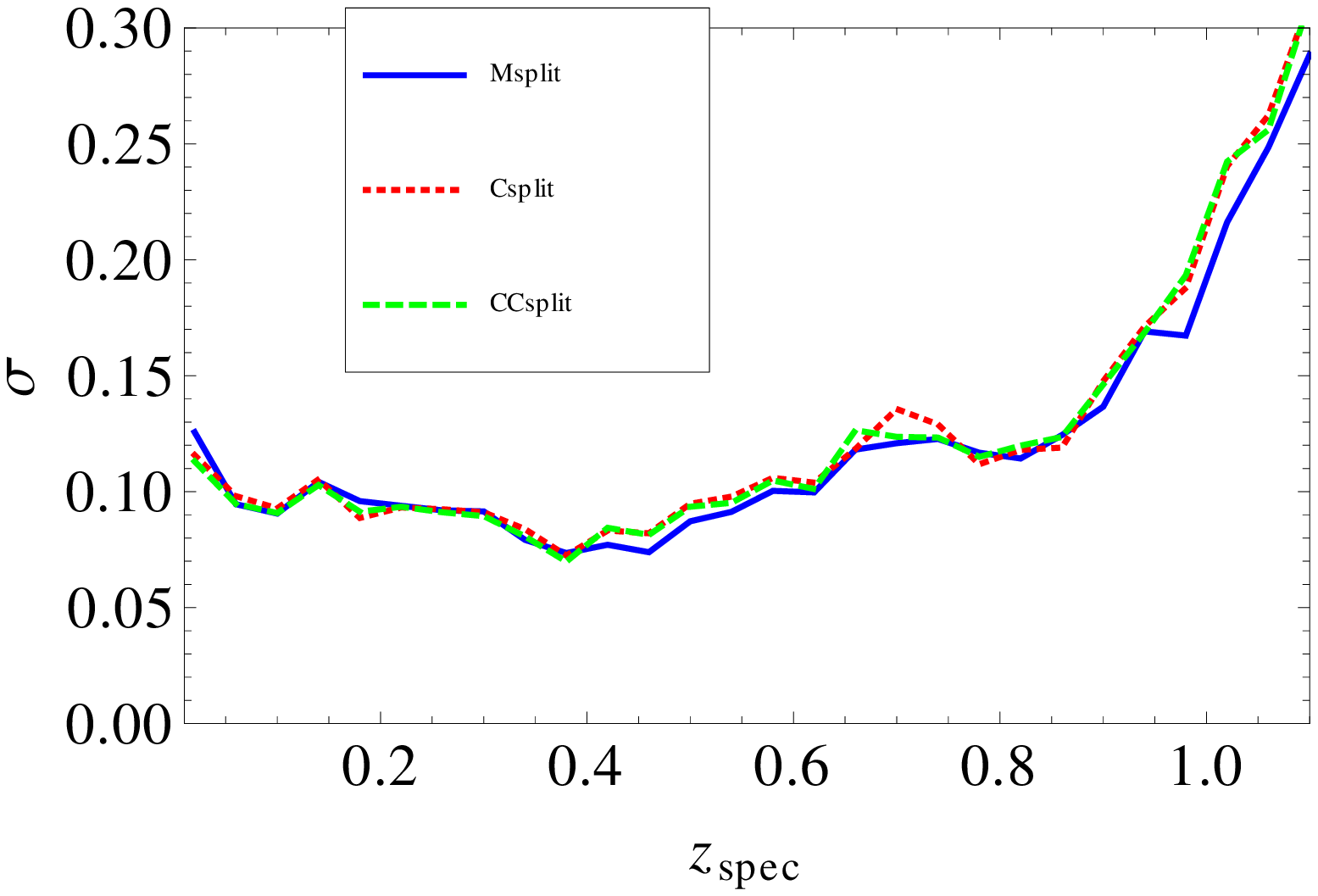}}
      \end{center}
    \end{minipage} \\
    \begin{minipage}[t]{80mm}
      \begin{center}
      \resizebox{80mm}{!}{\includegraphics[angle=0]{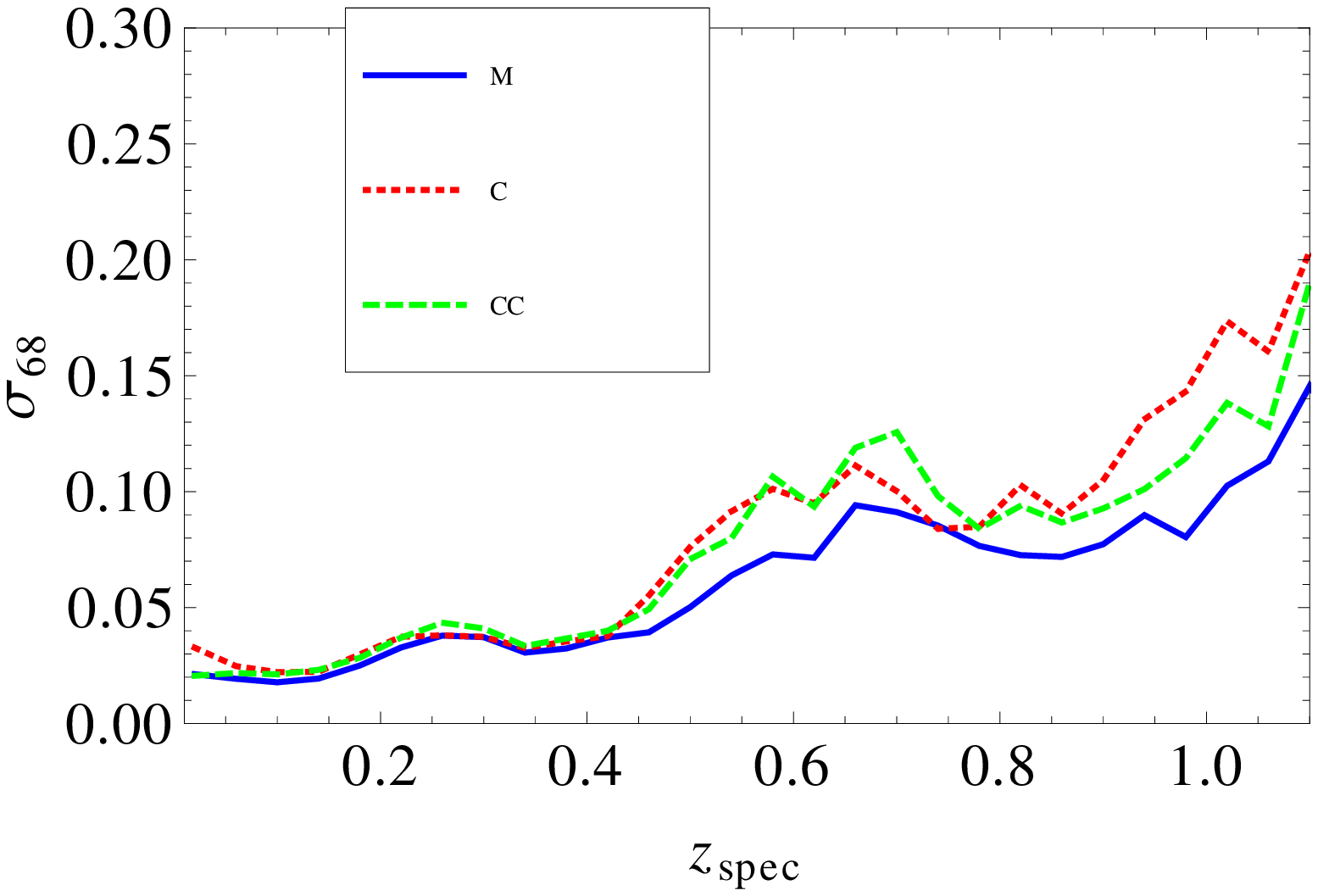}}
      \end{center}
    \end{minipage}
    \begin{minipage}[t]{80mm}
      \begin{center}
      \resizebox{80mm}{!}{\includegraphics[angle=0]{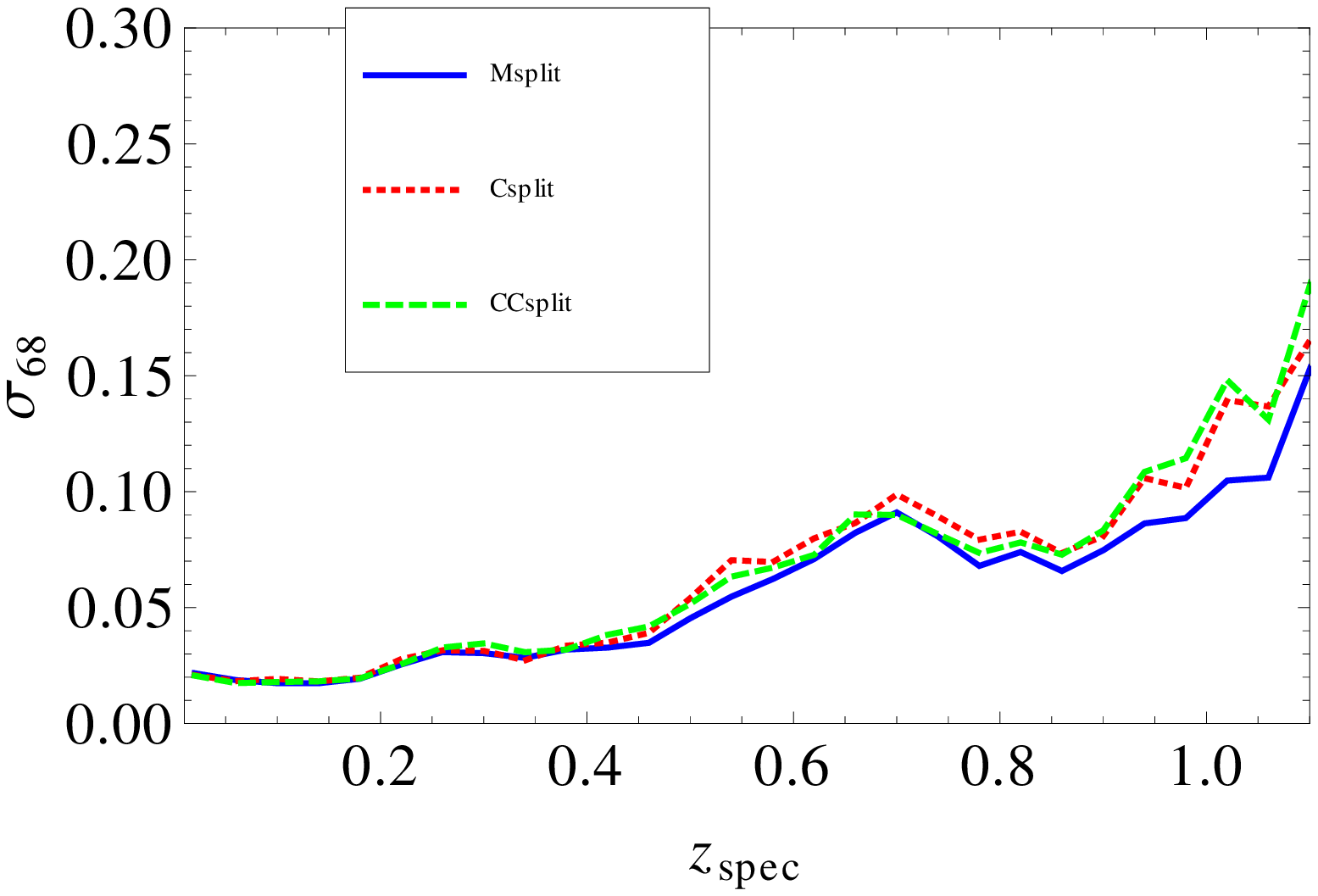}}
      \end{center}
    \end{minipage} 
  \end{center}
 \caption{$\sigma$ and $\sigma_{68}$ as a function of the spectroscopic redshift
 for all tested cases. 
 }
\label{sigma.vs.zspec}
\end{figure*}

Table \ref{table:sigma.summary} shows a summary of the performance results
of the different ANN cases. The standard deviation in this values, estimated from the five 
networks mentioned in Section \ref{subsec:meth_photoz}, is $0.001$. 
We also show in Figure \ref{sig:vs:mag} 
the performance indicators $\sigma$ and $\sigma_{68}$
as functions of $r$ magnitude for all cases. We see that the photo-z scatter increases
considerably for $r>22$. This effect can be explained by the small number of 
objects in the training set covering this regime (see Figure \ref{dist.spec}).
In addition, we show in Figs.
\ref{bias_vs_zphot} and \ref{sigma.vs.zphot} $z_{bias}$, $\sigma$ and $\sigma_{68}$
as functions of estimated photo-z and, in Figs. \ref{bias_vs_zspec} and \ref{sigma.vs.zspec},
the same indicators as functions of the the spectroscopic redshift. 
We can see that the values of these indicators increase for $z_{phot}>0.75$ 
regardless the case considered. 
We show in Table \ref{table:catast} another important indicator, the fraction of
catastrophic results, here defined as the number of objects for which we get
$|z_{phot}-z_{spec}|>0.1$ divided by the total number in the sample. This definition
corresponds to $\sim$ 12 \% of the distribution of $|z_{phot}-z_{spec}|$ for
this sample.
Based on theses results we choose 
Msplit as the best case. Specifically, Msplit has overall smaller $\sigma_{68}$ 
as a function of magnitude (Figure \ref{sig:vs:mag}) and a 
better fraction of catastrophic results (Table \ref{table:catast}).

\begin{figure}
  \begin{minipage}[t]{85mm}
    \begin{center}
      \resizebox{85mm}{!}{\includegraphics[angle=0]{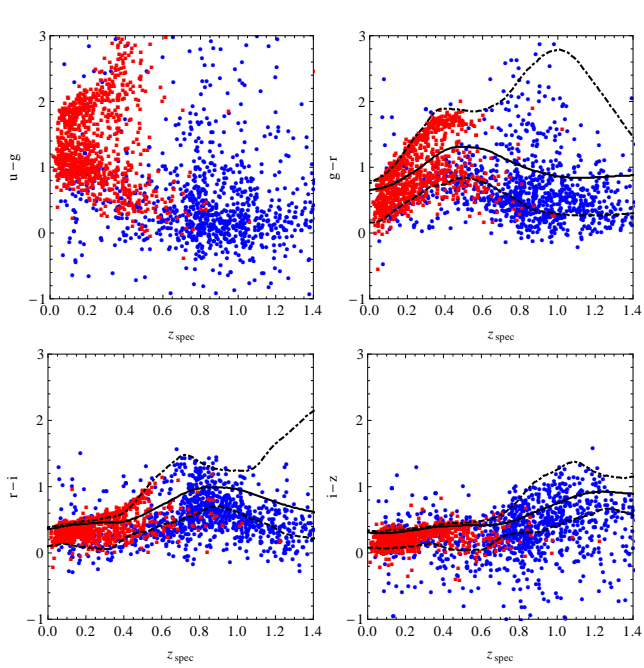}}
    \end{center}
  \end{minipage}
\caption{Colors vs spectroscopic redshift for galaxies in the validation set.
Red squares (blue circles) denote galaxies with $r<22$ ($r\geq22$). The curves 
are the predicted color-redshift relations for different types of galaxies 
(E,Sbc,Im) obtained by redshifting the k-corrected SEDs of 
\citet{assef2010} and applying the appropriate filters.
}
\label{color.zspec}
\end{figure}

In Fig. \ref{color.zspec} we plot the colors $u-g$, $g-r$, $r-i$ and $i-z$ 
versus spectroscopic redshift bright ($r<22$) and
faint ($r\geq 22$) galaxies in the validation set. We see that, for faint galaxies, 
colors and spectroscopic redshit are barely correlated. Such degeneracy
explains the low efficiency of the method in this magnitude regime.

\begin{figure}
  \begin{center}
    \begin{minipage}[t]{85mm}
      \begin{center}
      \resizebox{85mm}{!}{\includegraphics[angle=0]{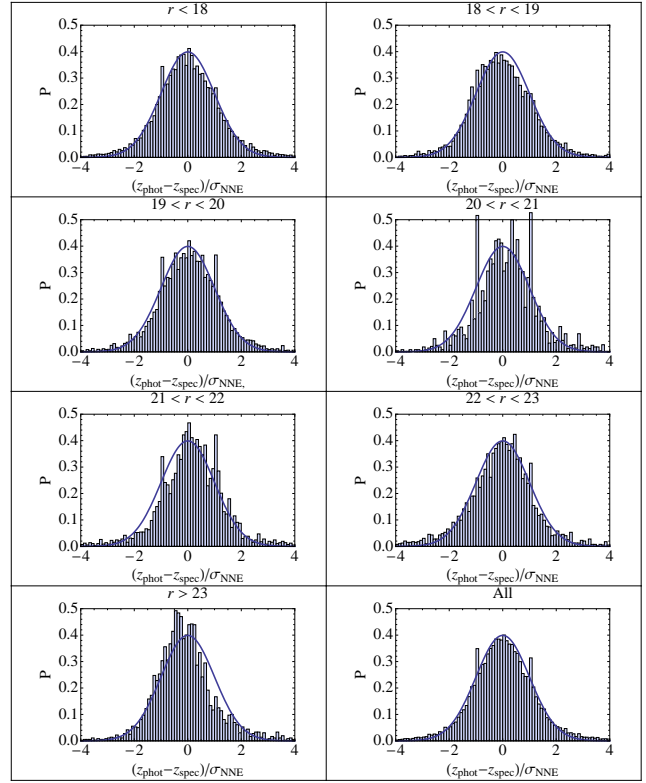}}
      \end{center}
    \end{minipage}
  \end{center}
 \caption{Distributions of $(z_{phot}-z_{spec})/\sigma_z^{NNE}$ for objects
 in the spectroscopic sample, in $r$ magnitude slices, for the Msplit case. 
 }
\label{errdist}
\end{figure}

In Fig. \ref{errdist} we plot the normalized error distribution, i.e., 
the distribution of $(z_{phot}-z_{spec})/\sigma_z^{NNE}$, for objects in 
the spectroscopic sample, using the Msplit case, in $r$ magnitude slices,
without any bias correction.
The solid lines show Gaussian distributions with zero mean and unit variance.
These plots indicate that, on average, the photo-z estimates are nearly
unbiased and the NNE error is a good estimate of the true error, although
we can see some asymmetry in the distribution depending on the magnitude 
range. 

\begin{figure}
  \begin{minipage}[t]{85mm}
    \begin{center}
      \resizebox{85mm}{!}{\includegraphics[angle=0]{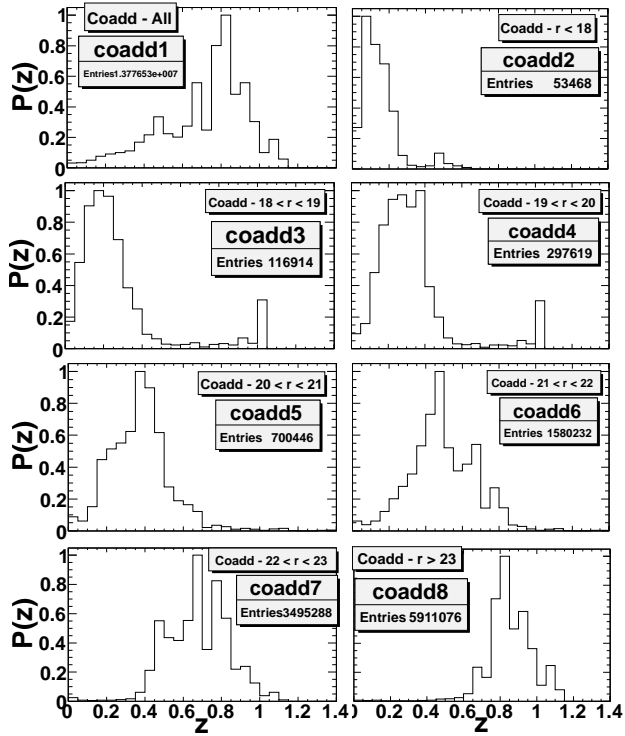}}
    \end{center}
  \end{minipage}
\caption{Photometric redshift distributions, corrected for bias, 
in $r$ magnitude slice for
the case Msplit.
}\label{coadd.magsplit.photodist}
\end{figure}

In Fig. \ref{coadd.magsplit.photodist}
we show the distribution of the estimated photometric redshift, corrected for the bias,
$z_{phot} - z_{bias}$ for the photometric sample,
in $r$ magnitude bins, for our best case (Msplit). The bias
was estimated from the validation sample in photo-z bins with
width $0.04$ as in Fig. \ref{bias_vs_zphot}. The bias correction
is included in the final catalog. 

\begin{figure}
  \begin{minipage}[t]{85mm}
    \begin{center}
      \resizebox{85mm}{!}{\includegraphics[angle=0]{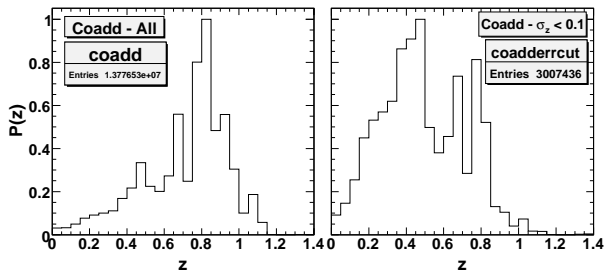}}
    \end{center}
  \end{minipage}
\caption{Photometric redshift distributions, for
the case Msplit. \textit{Left}: All objects. \textit{Right}: Objects with 
$\sigma_z^{NNE}<0.1$.
}\label{coadd.magsplit.photodist.cut}
\end{figure}

\begin{figure}
  \begin{center}
    \begin{minipage}[t]{50mm}
      \begin{center}
      \resizebox{50mm}{!}{\includegraphics[angle=0]{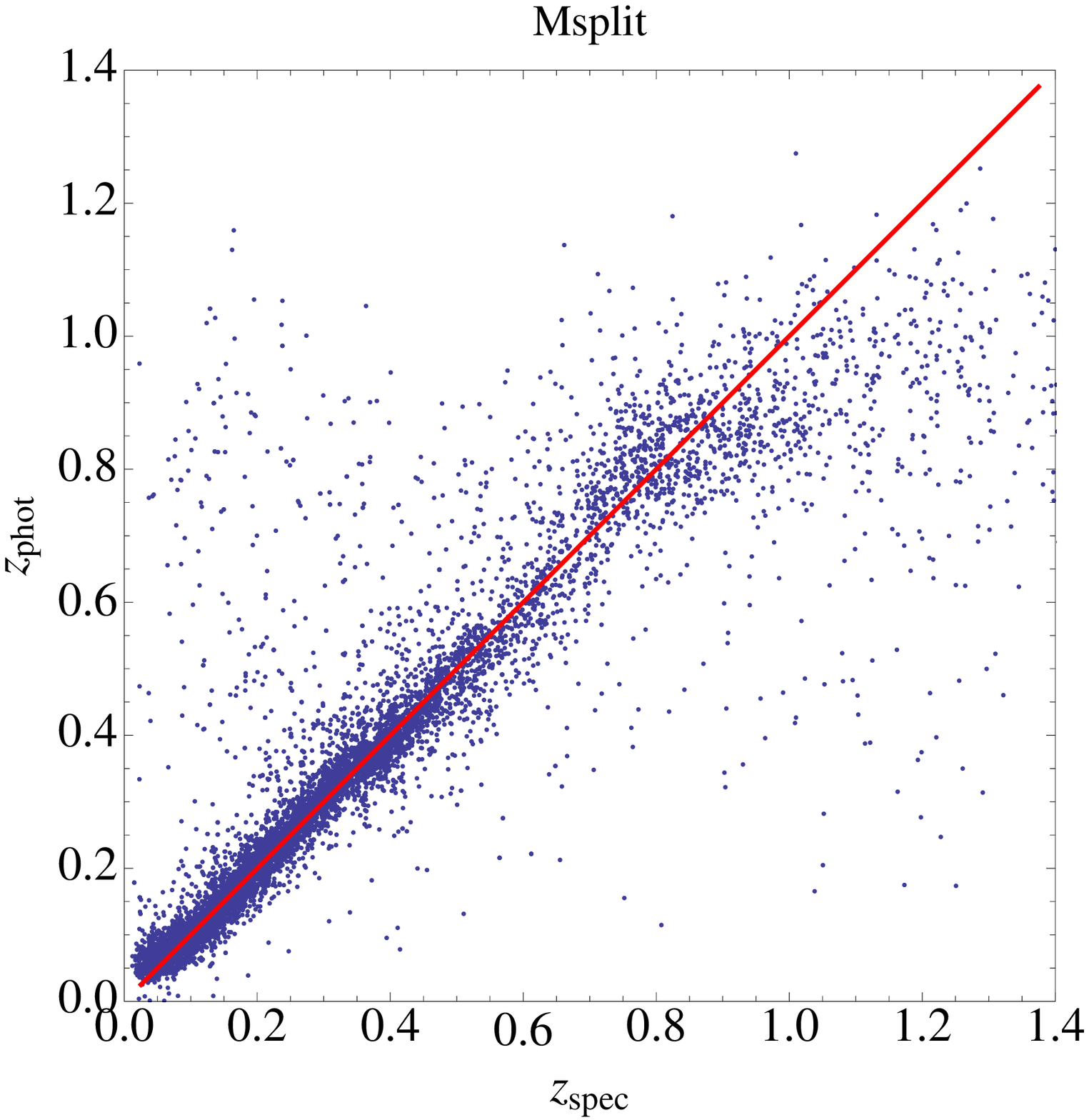}}
      \end{center}
    \end{minipage}
    \begin{minipage}[t]{50mm}
      \begin{center}
      \resizebox{50mm}{!}{\includegraphics[angle=0]{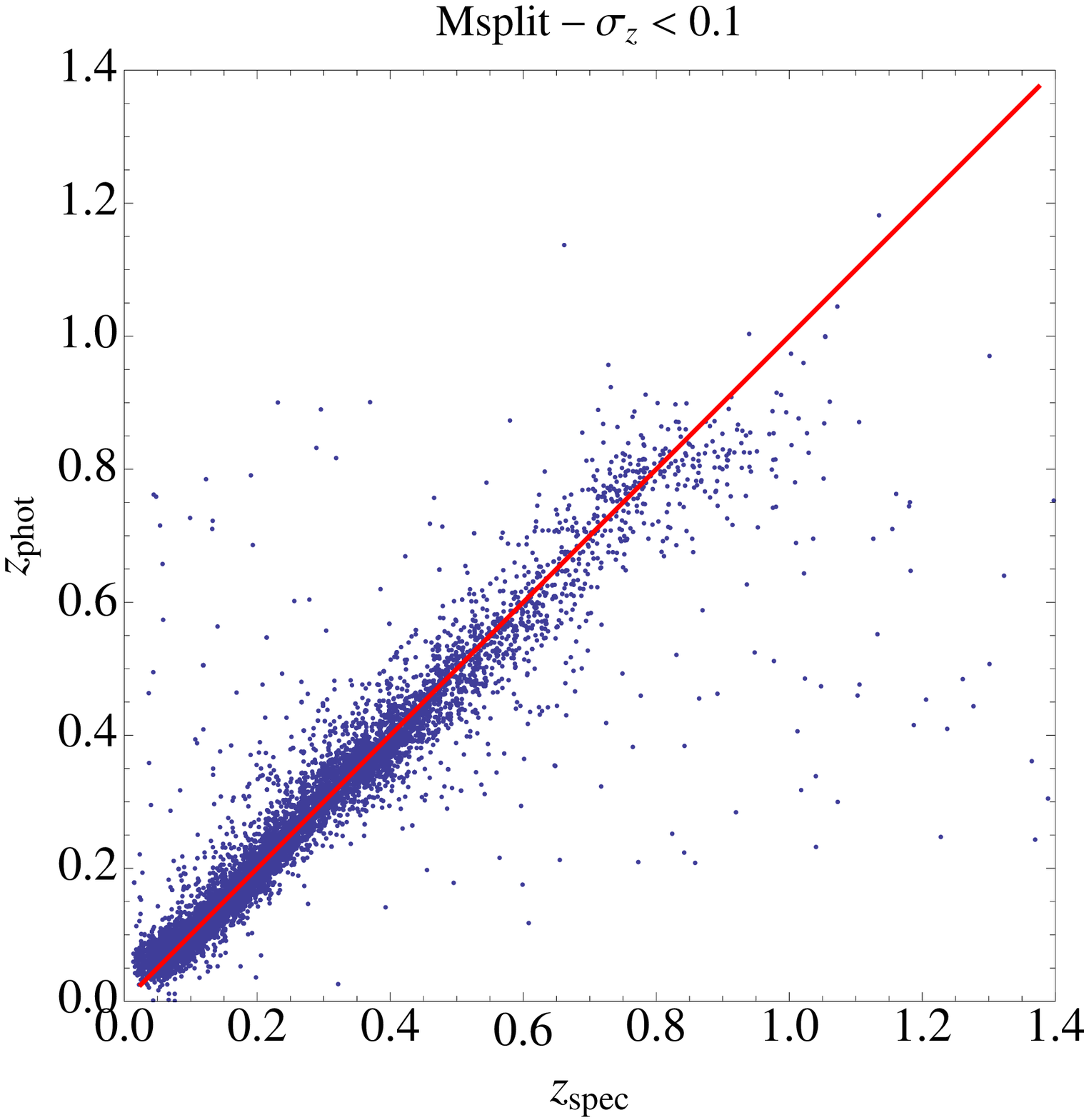}}
      \end{center}
    \end{minipage} 
  \end{center}
 \caption{Photo-z vs. spectroscopic redshift for Msplit case. \textit{Left}:
Full sample as in Fig. \ref{zpzs.valid_all}. \textit{Right}: Only objects with
$\sigma_z^{NNE}<0.1$. 
 }
\label{full_vs_errcut}
\end{figure}

For a significant fraction of the photometric sample the nearest neighbors
error estimate is large (greater than 10\% of photo-z value) and for most
of the science cases it will be necessary to cut the catalog. We show in 
Fig. \ref{coadd.magsplit.photodist.cut}, the photo-z distributions for
the whole sample (as in Fig. \ref{coadd.magsplit.photodist}) and for
objects with $\sigma_z^{NNE}<0.1$. We also show in Fig. \ref{full_vs_errcut}
the photometric redshift, $z_{phot}$,
for 10,000 randomly selected objects from validation set vs. true 
spectroscopic redshift, $z_{spec}$ for the same low error subsample.

We found that the use of concentration parameters does not improve
significantly the result, in contrast to our initial expectation,
based on the SDSS DR6 results \citep{oya08b}. \citet{omill11} also found
that these parameters improve the results for the SDSS DR7 main 
data. This is related to the error in the measured moments for higher magnitudes,
which is specially important for this sample, consequently 
the additional noise roughly compensates 
the additional information from these parameters. Similar conclusions
can be found in \citep{sin11}, although their definition of concentration
is not the same used here.

\begin{deluxetable}{ll}
\tablewidth{0pt}
\tablecaption{Description of the different combinations}
\startdata
\hline
\hline
\multicolumn{1}{c}{Case} & \multicolumn{1}{c}{Inputs/Description} \\
\hline
C & $u-g$, $g-r$, $r-i$, $i-z$                  \\  
Csplit & $u-g$, $g-r$, $r-i$, $i-z$, split in $r$ slices      \\  
M & $u$, $g$, $r$, $i$, $z$                \\  
Msplit & $u$, $g$, $r$, $i$, $z$, split in $r$ slices                \\ 
CC &$u-g$, $g-r$, $r-i$, $i-z$ + $c_u,c_g,c_r,c_i,c_z$   \\
CCsplit &$u-g$, $g-r$, $r-i$, $i-z$ + $c_u,c_g,c_r,c_i,c_z$, split in $r$ slices   \\
\enddata
\label{table:ANNcases}
\end{deluxetable}

\begin{deluxetable}{lcc}
\tablewidth{0pt}
\tablecaption{Summary of ANN cases}
\startdata
\hline
\hline
\multicolumn{1}{c}{Case} &  \multicolumn{1}{c}{$\sigma$} & \multicolumn{1}{c}{$\sigma_{68}$}\\
\hline
C                  & 0.16 & 0.046 \\  
Csplit    & 0.14 & 0.034\\
M    & 0.14 & 0.034\\
Msplit    & 0.14 & 0.031\\
CC                  & 0.15 & 0.043 \\  
CCsplit    & 0.14 & 0.032\\
\enddata
\label{table:sigma.summary}
\tablecomments{$\sigma$ and $\sigma_{68}$ 
for the validation set using different input parameters 
(magnitudes, colors, and concentration indices) and training procedures
(training with the whole sample or in magnitude bins independently).}
\end{deluxetable}
\pagebreak

\begin{deluxetable}{l|c|c|c|c|c|c|c|c}
\tablewidth{0pt}
\tablecaption{Catastrophic redshifts}
\startdata
\hline
\hline
\backslashbox{Case}{Range}  & $r<18$
& $18<r<19$ & $19<r<20$ & $20<r<21$ & $21<r<22$ & $22<r<23$ & $r>23$  & all \\
\hline
C    & 0.020  & 0.034  & 0.048  & 0.092 & 0.14  & 0.22  & 0.17  & 0.075 \\  
Csplit   & 0.0013  & 0.0063  & 0.0058  & 0.093  & 0.084  & 0.28  & 0.29  & 0.062 \\  
M    & 0.0012  & 0.0034  & 0.012  & 0.054  & 0.10  & 0.26  & 0.26  & 0.058  \\  
Msplit   & 0.0012  & 0.0042  & 0.0068  & 0.059  & 0.11  & 0.25  & 0.24  & 0.055  \\
CC   & 0.013  & 0.022  & 0.030  & 0.066  & 0.13  & 0.25  & 0.21  & 0.069  \\
CCsplit   & 0.0012  & 0.0053  & 0.0056  & 0.089  & 0.083  & 0.28  & 0.28  & 0.060  \\
\enddata
\label{table:catast}
\tablecomments{Fraction of objects ($N_{cat}/N_{total}$) with $|z_{phot}-z_{spec}|>0.1$
for the validation set using different input parameters 
(colors, concentration indices and magnitudes) and training procedures.}
\end{deluxetable}

\section{Accessing the Catalog} \label{cat}

The best case bias corrected photo-z catalog (Msplit) is publicly available
as a SDSS value-added catalog at 
{\tt http://www.sdss.org/dr7/products/value\_ added/index.html}.

\section{Conclusions}\label{con}

We have presented a public catalog of photometric redshifts for the SDSS coadd 
photometric sample using  
photo-z estimates, based on the ANN method, considering the five
magnitudes $ugriz$ as input parameters and also performing the
training in $r$ magnitude bins separately (Msplit). Our tests indicate 
that the photo-z estimates are most reliable for galaxies 
with $r<22$
and that the scatter increases significantly at fainter magnitudes.
Based on our results, we advise the reader to use carefully this catalog
for $z_{phot}\geq 0.75$, since all performance indicators show a
lower efficiency of the method, with the chosen spectroscopic 
sample, at this redshift range. However, depending on the specific science goals, a simple quality 
cut on the photo-z error might be
sufficient to compensate this problem at the desired level.  

\acknowledgments

Funding for the Sloan Digital Sky Survey (SDSS) and SDSS-II has been 
provided by the Alfred P. Sloan Foundation, the Participating Institutions, 
the National Science Foundation, the U.S. Department of Energy, the National 
Aeronautics and Space Administration, the Japanese Monbukagakusho, and the Max 
Planck Society, and the Higher Education Funding Council for England. 
The SDSS Web site is {\tt http://www.sdss.org/}.

The SDSS is managed by the Astrophysical Research Consortium (ARC) for the 
Participating Institutions. The Participating Institutions are the American 
Museum of Natural History, Astrophysical Institute Potsdam, University of Basel, 
University of Cambridge, Case Western Reserve University, The University of 
Chicago, Drexel University, Fermilab, the Institute for Advanced Study, the 
Japan Participation Group, The Johns Hopkins University, the Joint Institute 
for Nuclear Astrophysics, the Kavli Institute for Particle Astrophysics and Cosmology, 
the Korean Scientist Group, the Chinese Academy of Sciences (LAMOST), Los Alamos 
National Laboratory, the Max-Planck-Institute for Astronomy (MPIA), the Max-Planck-Institute 
for Astrophysics (MPA), New Mexico State University, Ohio State University, University of 
Pittsburgh, University of Portsmouth, Princeton University, the United States Naval 
Observatory, and the University of Washington.

\appendix

\section{Data Query Code}\label{query}

Here we provide the SDSS database query used to obtain the catalog containing  
the photometric sample used in this paper. 
Notice that the query requires the TYPE flag to be set to 3 (galaxies) and 
selects objects with dereddened model magnitude  $16<r<24.5$, 
which do not have any of the following flags:
\verb1BRIGHT1, \verb1SATURATED1 and \verb1SATUR_CENTER1. 
The full query is shown below

\vspace{0.8 cm}

\begin{verbatim}
SELECT  ObjID,ra,dec,
        dered_u,dered_g,dered_r,dered_i,dered_z,
        petroR50_u/petroR90_u as c_u,petroR50_g/petroR90_g as c_g,
        petroR50_r/petroR90_r as c_r,
        petroR50_i/petroR90_i as c_i,petroR50_z/petroR90_z as c_z,
        err_u,err_g,err_r,err_i,err_z
INTO coadd_mags_allinone
FROM Stripe82..PhotoObjAll
WHERE (flags_r & 0x0000080000040002)=0
        AND type=3
        AND mode=1
        AND (run=106 or run=206)
        AND dered_r BETWEEN 16 AND 24.5
\end{verbatim}
We made an additional cut in order to select only objects which have positive values for
petroR50/petroR90. The final catalog has 13,688,828 galaxies.

\vspace{0.5cm}

Here we provide a brief description of the flags used in the query:
BRIGHT indicates that an object is a duplicate detection of an object with 
signal to noise greater  
than $200 \sigma$; SATURATED indicates that an 
object contains one or more saturated pixels; 
SATUR\_CENTER indicates that the object center is close to at least one 
saturated pixel.
Note that in selecting PRIMARY objects (using PhotoPrimary), 
we have implicitly selected objects 
that either do {\it not} have the BLENDED flag set 
or else have NODEBLEND set or nchild equal zero. 
In addition, the PRIMARY catalog contains no BRIGHT objects, so 
the cut on BRIGHT objects in the query above is in fact redundant. 
BLENDED objects have multiple peaks detected within them, which PHOTO  
attempts to deblend into several CHILD objects. 
NODEBLEND objects are BLENDED but no deblending was attempted on them, because
they are either too close to an EDGE, or too large, or one of 
their children overlaps an edge. A few percent of the objects in 
our photometric sample have NODEBLEND set; some users may wish to 
remove them.

We also suggest that users require objects to have the 
BINNED1 flag set. 
BINNED1 objects were detected at $\geq 5 \sigma$ significance 
in the original imaging frame.

The SDSS webpage
\footnote{\tt{http://cas.sdss.org/dr7/en/help/docs/algorithm.asp?search=flags\&submit1=Search}} provides
further recommendations about flags, which we strongly recommend that users read.

\bibliographystyle{apj}
\bibliography{coadd_photoz_after_ref}

\end{document}